\documentclass[lettersize,journal]{IEEEtran}
\usepackage{amsmath,amsfonts}
\usepackage{algorithm}
\usepackage{array}
\usepackage[caption=false,font=normalsize,labelfont=sf,textfont=sf]{subfig}
\usepackage{textcomp}
\usepackage{stfloats}
\usepackage{url}
\usepackage{verbatim}
\usepackage{graphicx}
\usepackage{cite}
\usepackage{algpseudocode}
\usepackage{hyperref}
\usepackage{booktabs}
\usepackage{extra_styles}

\begin{document}

\title{EDNet: A Versatile Speech Enhancement Framework with Gating Mamba Mechanism and Phase Shift-Invariant Training}

\author{Doyeop Kwak, Youngjoon Jang, Seongyu Kim, and Joon Son Chung
\thanks{The authors are with the School of Electrical Engineering, Korea Advanced Institute of Science and Technology, Daejeon 34141, Republic of Korea (e-mail: dobbyk@kaist.ac.kr; wgs01088@kaist.ac.kr; kimsk46@kaist.ac.kr; joonson@kaist.ac.kr)}%
\thanks{A preprint version of this work is available on arXiv:2506.16231. \cite{kwak2025ednetdistortionagnosticspeechenhancement}}
}

\markboth{Journal of \LaTeX\ Class Files,~Vol.~14, No.~8, August~2021}%
{Shell \MakeLowercase{\textit{et al.}}: A Sample Article Using IEEEtran.cls for IEEE Journals}

\maketitle

\begin{abstract}
Speech signals in real-world environments are frequently affected by various distortions such as additive noise, reverberation, and bandwidth limitation, which may appear individually or in combination. Traditional speech enhancement methods typically rely on either masking, which focuses on suppressing non-speech components while preserving observable structure, or mapping, which seeks to recover clean speech through direct transformation of the input. Each approach offers strengths in specific scenarios but may be less effective outside its target conditions. We propose the Erase and Draw Network (EDNet), a versatile speech enhancement framework designed to handle a broad range of distortion types without prior assumptions about task or input characteristics. EDNet consists of two main components: (1) the Gating Mamba (GM) module, which adaptively combines masking and mapping through a learnable gating mechanism that selects between suppression (Erase) and reconstruction (Draw) based on local signal features, and (2) Phase Shift-Invariant Training (PSIT), a shift tolerant supervision strategy that improves phase estimation by enabling dynamic alignment during training while remaining compatible with standard loss functions. Experimental results on denoising, dereverberation, bandwidth extension, and multi distortion enhancement tasks show that EDNet consistently achieves strong performance across conditions, demonstrating its architectural flexibility and adaptability to diverse task settings.

\end{abstract}


\section{Introduction}

\IEEEPARstart{I}{n} real-world conditions, speech signals are frequently corrupted by diverse acoustic distortions such as additive noise, reverberation, and bandwidth limitations. These degradations often co-occur in unpredictable combinations, severely hindering both perceptual intelligibility and the performance of downstream applications~\cite{weninger2015speech,desjardins2014effect, kwak2024voxmm}. While speech enhancement (SE) systems essentially aim to restore clean speech from such degraded inputs, most existing approaches are optimized for a specific type of distortion. This task-specific design paradigm may fail to reflect the reality of practical deployment environments, where distortion types are neither known in advance nor mutually exclusive, and may even vary over time.

Conventional SE methods can be largely categorized into masking-based and mapping-based approaches. Masking-based methods suppress non-speech components while preserving speech structure by predicting a mask, making them effective for tasks such as denoising~\cite{ pascual2017segan, yin2022tridentse, hu2001speech, yilmaz2004blind, narayanan2013ideal, srinivasan2006binary, lu2023explicit, chao2024investigation} and dereverberation~\cite{roman2011intelligibility, may2014generalization, li2017ideal, kothapally2022skipconvgan, williamson2017time}. However, their reliance on input spectral cues limits their ability to reconstruct fully lost speech components~\cite{chao2025universal}. In contrast, mapping-based methods directly transform degraded speech into clean speech, facilitating the reconstruction of missing components—especially in tasks such as bandwidth extension~\cite{kuleshov2017audio, abel2018simple, li2015deep, wang2021towards, lu2024towards, liu2022neural}. Despite these strengths, these methods can excessively alter preserved speech content, undermining identity preservation and leading to suboptimal objective scores~\cite{nossier2020mapping, kumar2020nu, liu2022neural}. These contrasting behaviors highlight the inherent trade-off between preservation and reconstruction in existing SE strategies. Several hybrid approaches~\cite{zhang2021deep, zhang2023dual, hao2020masking, abdulatif2024cmgan} have been proposed to leverage the complementary strengths of masking and mapping, aiming to balance selective suppression with generative reconstruction. These methods have shown effectiveness when tailored to specific enhancement tasks, but most integrate the two methods through fixed fusion schemes—such as summation or averaging—that are agnostic to the characteristics of the input. While such strategies can work well in certain cases, applying a static combination regardless of distortion type or task may fall short of optimality and often requires manual design choices about how the fusion should be performed.


This work explores whether a model can adapt its processing strategy to the characteristics of input distortions through architectural design, without relying on task-specific assumptions or explicit supervision. Ultimately, our goal is to embed inductive biases that enable the model to identify and apply the most effective processing strategy, even under complex conditions where multiple distortions interact and the optimal pathway is not easily defined. To this end, we present Erase and Draw Network (EDNet), a speech enhancement framework designed to accommodate diverse and compound distortions through architectural flexibility. The central component of EDNet is the Gating Mamba (GM) module, which adaptively integrates masking and mapping strategies for magnitude refinement based on input characteristics. Unlike prior hybrid methods that statically fuse both outputs, the GM module employs a learnable gating function to determine, on a region-wise basis, whether to suppress (Erase) non-speech components or to reconstruct (Draw) missing information. This enables the model to dynamically adjust its enhancement behavior, modulating the balance between structure preservation and content generation according to the nature of the distortion. By embedding this form of control into the model itself, EDNet provides a unified architecture that can adapt to a wide range of distortions and tasks, without relying on task-specific designs or prior knowledge of the distortion type.

Complementing this architectural flexibility on the magnitude side, we also address the orthogonal challenge of phase reconstruction. Accurate phase estimation is crucial for perceptual quality, particularly in conditions involving reverberation or multiple interacting distortions~\cite{paliwal2011importance}. However, phase learning is inherently difficult due to its periodicity and sensitivity to temporal shifts~\cite{masuyama2020phase}. Conventional phase reconstruction methods~\cite{takamichi2018phase,ai2023neural,zhang2024unrestricted,masuyama2020phase} typically constrain training to a single ground truth (GT) phase per sample, implicitly requiring the model to align its output exactly with the GT phase. This strict constraint can be unnecessary, as minor phase shifts are, from a signal processing perspective, nearly indistinguishable from small temporal shifts in the time domain, and therefore have negligible impact on perceived speech quality~\cite{ku2024explicit}.
While some prior works~\cite{lee2023phaseaug, ku2024explicit} attempt to address these issues through data augmentation or specialized objectives, such methods often sacrifice compatibility with standard loss functions or require non-standard training pipelines.

To address this, we propose Phase Shift-Invariant Training (PSIT), a simple yet effective strategy that reframes phase prediction as a shift-tolerant one-to-many problem. PSIT dynamically applies an optimal alignment during training, relaxing the strict alignment constraint without altering standard loss functions or pipelines. By reducing the excessive penalties caused by misaligned phase errors, PSIT facilitates more effective and focused learning of the phase structural refinement itself. Moreover, it mitigates the risk of incoherent gradients propagating from the phase branch into the magnitude branch, which could otherwise destabilize magnitude optimization and degrade overall training efficiency. 
In addition to proposing the PSIT methodology, we conduct a series of analyses to better understand the impact of shift alignment constraints and their relaxation in terms of speech quality metrics, phase reconstruction performance, and overall training dynamics. These findings may offer practical insight for future studies on phase modeling in speech processing.

We evaluate EDNet on three representative speech enhancement benchmarks—denoising, dereverberation, and bandwidth extension—and observe state-of-the-art or competitive performance across all tasks. To further assess structural adaptability and phase modeling capabilities, we conduct two additional experiments: multi-distortion enhancement, where EDNet outperforms prior models under compounded distortions, and phase reconstruction, where PSIT yields statistically significant gains in phase estimation accuracy. Our key contributions are summarized as follows:

\begin{itemize}
\item We propose EDNet, a versatile speech enhancement framework designed to handle diverse and compound distortions by embedding structural adaptability into its architecture.
\item We introduce the Gating Mamba (GM) module, which adaptively integrates masking and mapping operations for flexible control between suppression and reconstruction.
\item We present Phase Shift-Invariant Training (PSIT), a shift-tolerant phase supervision strategy that improves phase recovery and training efficiency while maintaining compatibility with standard loss functions.
\item Extensive experimental analysis across single- and multi-distortion tasks validate the effectiveness of both EDNet’s architectural adaptability and its training strategy.
\end{itemize}
\section{Related Work}
\label{sec:related_work}

\subsection{Masking-based SE}

Masking-based methods operate primarily in the time-frequency (TF) domain by learning masks that selectively suppress non-speech components while preserving speech-dominant regions. Early works like the Ideal Binary Mask (IBM)~\cite{hu2001speech, yilmaz2004blind, roman2011intelligibility, may2014generalization} apply hard decisions to each TF unit, significantly improving intelligibility but introducing artifacts due to coarse quantization~\cite{hummersone2014ideal, jin2009supervised}. To address this, Ideal Ratio Mask (IRM)~\cite{narayanan2013ideal, srinivasan2006binary, li2017ideal}  was proposed, enabling soft masks with continuous values between 0 and 1 for smoother enhancement. The Spectral Magnitude Mask (SMM)~\cite{wang2014training}  further relaxes the range constraint, allowing unbounded mask values and increasing modeling flexibility. As phase information gained importance, methods such as Phase-Sensitive Mask (PSM)~\cite{weninger2015speech,erdogan2015phase}  and Complex Ideal Ratio Mask (cIRM)~\cite{williamson2015complex, kothapally2022skipconvgan, williamson2017time}  were developed. Masking has been also extended to the latent feature space in time-domain architectures such as TasNet~\cite{luo2018tasnet}, where masks are applied directly to learned feature representations, avoiding the constraints of the STFT domain. This feature-level masking has become central in many state-of-the-art models, particularly in the context of speech separation~\cite{subakan2021attention, zhao2023mossformer}. 
Despite their wide adoption, masking-based models remain fundamentally limited in scenarios involving severely missing spectral information, such as bandwidth limitation and packet loss~\cite{chao2025universal}. 

\subsection{Mapping-based SE}

Mapping-based speech enhancement methods aim to learn a nonlinear transformation that reconstructs clean speech directly from noisy input, typically in the waveform or spectral domain. Although masking remains the dominant approach in denoising, several studies have explored mapping-based alternatives. A representative example is SEGAN~\cite{pascual2017segan}, which employs adversarial training for waveform mapping. Subsequent works increased generator capacity~\cite{phan2020improving} and combined adversarial loss with spectral-domain losses~\cite{pascual2019towards} for improved stability and fidelity. In dereverberation, mapping-based methods are motivated by the need of modeling long-range temporal smearing introduced by room reverberation. Early spectral regression models~\cite{han2015learning} used fully connected networks to predict clean spectra, followed by U-Net structures~\cite{ernst2018speech}, and variants that replaced skip connections with multiple convolutional modules to improve feature propagation~\cite{kothapally2020skipconvnet}. More recently, deformable convolutional networks (DCN)~\cite{kothapally2024monaural} have been introduced context-sensitive transformations that extend beyond conventional masking and exhibit characteristics of both filtering and reconstruction-based approaches. Bandwidth extension is an area where mapping dominates, as it requires inferring missing high-frequency content. TF-domain methods~\cite{abel2018simple, li2015deep, lu2024towards} estimate high-band magnitude spectra from narrowband inputs and use heuristics or learned strategies for phase reconstruction. Time-domain approaches~\cite{kuleshov2017audio, wang2021towards} directly predict wideband waveforms, while vododer-based method such as NVSR~\cite{liu2022neural} uses a two-stage pipeline of spectral mapping followed by waveform generation via a neural vocoder. Despite their generative flexibility, mapping-based models can inadvertently distort clean input regions due to their global transformation behavior~\cite{nossier2020mapping, kumar2020nu, liu2022neural}. 

\subsection{Hybrid Approachs}

To overcome the individual limitations of masking and mapping approaches, several studies have explored hybrid architectures that combine the two, aiming to take advantage of their complementary strengths: masking offers selective noise suppression, while mapping enables detailed signal reconstruction. PhaseDCN~\cite{zhang2021deep} applies the hybrid approach with a focus on improving phase modeling to enhance denoising performance. It first generates a masked magnitude spectrum using an IRM and concatenates it with the noisy input. The combined features are fed into a complex-valued mapping network that produces a final output, implementing early stage fusion at the feature level. DBDIUNet~\cite{zhang2023dual} focuses on noisy and reverberant conditions by predicting a cIRM and a magnitude spectrogram through dual decoders, then averaging their outputs to produce the final result. The masking-and-inpainting~\cite{hao2020masking} method targets low signal-to-noise ratios (SNR) and non-stationary noise by first removing highly corrupted time-frequency regions using a binary mask, then reconstructing the removed region with a convolutional inpainting module. These two stages are executed sequentially, without an explicit fusion step. CMGAN~\cite{abdulatif2024cmgan}, designed to handle various SE tasks such as denoising, dereverberation, and bandwidth extension, predicts magnitude masks and complex spectrograms in parallel and combines them using element-wise summation. Similarly, HD-DEMUCS~\cite{kim2023hd} targets general speech restoration using two parallel heterogeneous decoders, one for suppression and one for refinement, which are combined at the waveform level using a learnable weighting factor. This demonstrates the potential of hybrid methods in building general speech enhancement systems.

Despite differences in design and application, these approaches share several limitations. Most rely on fixed fusion strategies—such as averaging, concatenation, or summation—that are agnostic to input characteristics. While methods such as HD-DEMUCS introduce learnable fusion, the weighting mechanism operates only on the final decoder outputs, without direct guidance from the original input features. Moreover, these fusion mechanisms are typically applied globally at the output stage, lacking the ability to modulate the region-wise contributions of masking and mapping dynamically across time and frequency.
These observations underscore the need for more flexible and sophisticated integration strategies that can coordinate masking and mapping in a input content-aware, task-agnostic manner.

\subsection{Phase Modeling}

Early speech enhancement models primarily focused on magnitude estimation~\cite{xu2014regression, ai2019dnn}, often reusing the noisy phase during synthesis. However, it is now widely recognized that accurate phase reconstruction is essential for achieving high perceptual quality~\cite{paliwal2011importance}. One line of research addresses phase prediction by directly operating in the complex domain either through masking~\cite{hao2021fullsubnet,choi2018phase, zhao2022frcrn} or mapping~\cite{wang2023tf,wang2020complex}, where both magnitude and phase are implicitly modeled via real and imaginary components. While conceptually straightforward, this approach often suffers from optimization instability and limited interpretability, especially when paired with magnitude-focused objectives such as spectral magnitude loss or mask estimation targets~\cite{lu2023explicit}. To improve the stability and effectiveness of phase modeling, recent studies~\cite{chao2024investigation, lu2023explicit, lu2024towards} have moved toward explicitly decoupling magnitude and phase and modeling them through separate branches.

Supervision for phase modeling varies by modeling approach. In complex-valued formulations, losses are typically applied in terms of complex spectral distance. In contrast, explicit phase prediction often requires trigonometric losses~\cite{takamichi2018phase} or anti-wrapping functions~\cite{ai2023neural} to handle phase discontinuities, and is frequently combined with derivative-based objectives such as group delay (GD) or instantaneous angular frequency (IAF)~\cite{zhang2024unrestricted, masuyama2020phase}. Most of these methods, however, assume a single ground-truth (GT) phase, implicitly enforcing strict alignment, which can penalize perceptually irrelevant shifts and destabilize learning~\cite{ku2024explicit}.

While the limitations of strict phase alignment have begun to attract attention, only a few studies have explored ways to relax this constraint. PhaseAug~\cite{lee2023phaseaug}, originally proposed for vocoder training, introduces random phase shifts during training to expose the model to multiple plausible alignments. However, the loss is still computed with respect to a single shifted target, meaning the model is implicitly encouraged to align with the shifted version of the GT.
Ku et al.~\cite{ku2024explicit} take a different approach by removing phase GT entirely and using magnitude-phase consistency as an indirect supervision signal. While this avoids explicit alignment, it prevents the use of standard loss functions—such as waveform or complex spectral losses—since applying them would implicitly reintroduce supervision toward a fixed target phase.
These issues suggest the need for a supervision strategy that resolves alignment ambiguity while preserving compatibility with existing loss functions and training pipelines.

\begin{figure*}[!t]
   \centering
   \vspace{-3mm}
   \includegraphics[width=0.99\linewidth]{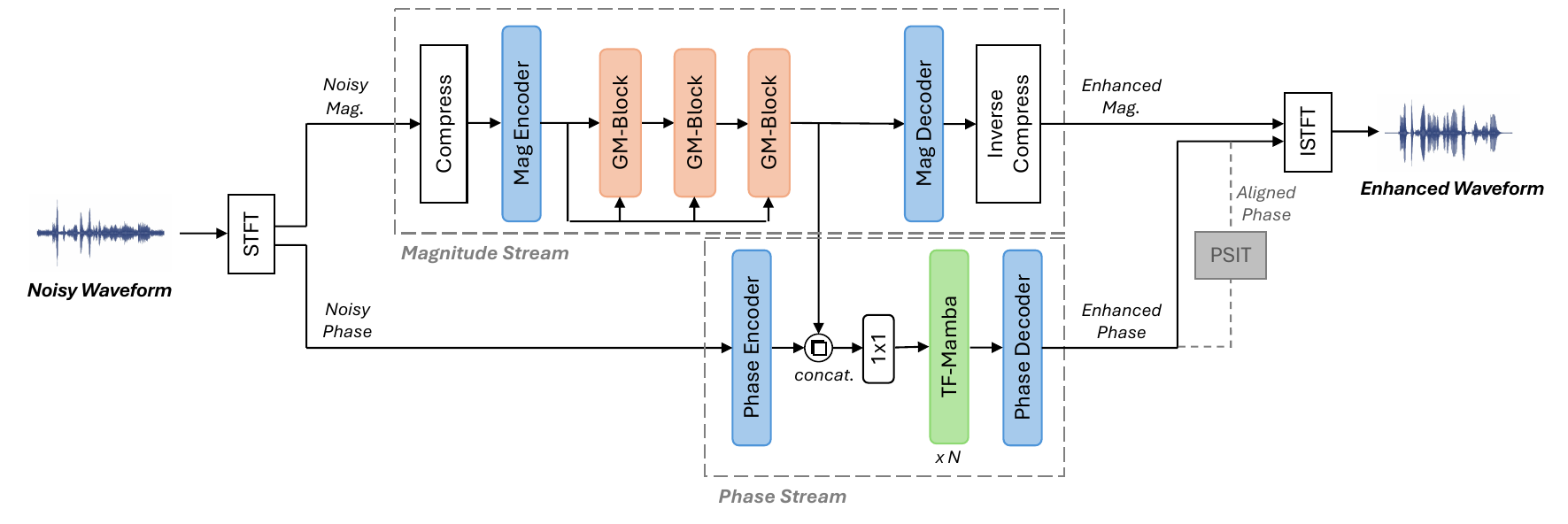}
   \vspace{-4mm}
   \caption{Overview of EDNet architecture. The model employs separate magnitude and phase enhancement streams to handle their distinct characteristics. PSIT is applied only during training to improve phase reconstruction without affecting the inference pipeline.}
   \label{fig:arch}
   \vspace{-3mm}
\end{figure*}

\begin{figure}[t]
   \centering
   \vspace{-2mm}
   \includegraphics[width=0.99\linewidth]{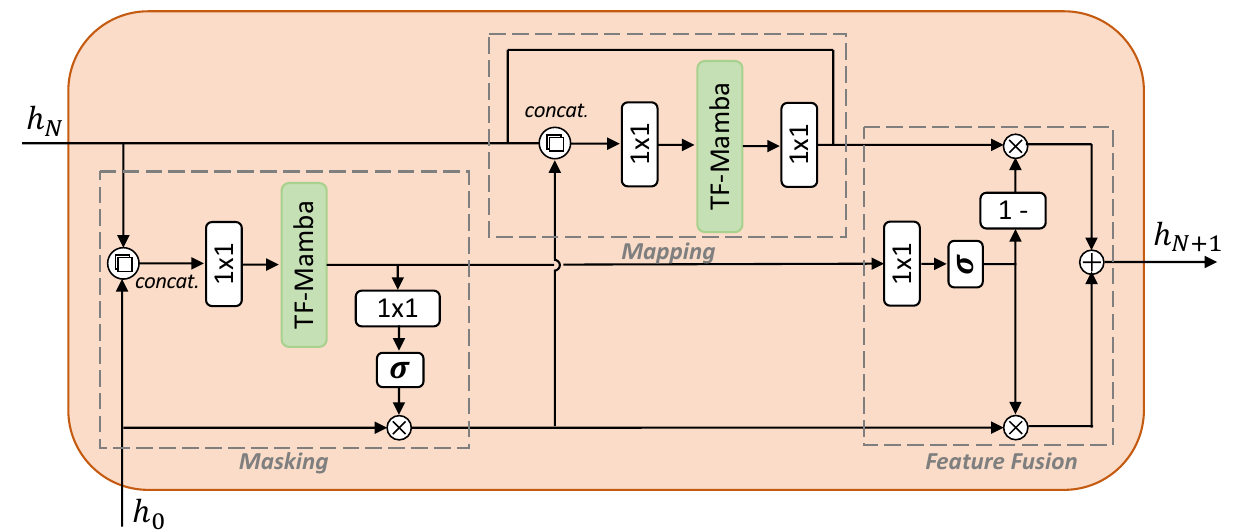}
   \vspace{-2mm}
   \caption{Detailed architecture of the GM block. $h_N$ represents the hidden feature from the previous block, while $h_0$ refer to the initial hidden feature from the encoder.}
   \label{fig:gm_module}
\vspace{-4mm}
\end{figure}
\section{Methodology}
As discussed in~\Sref{sec:related_work}, the effectiveness of existing speech enhancement methods is often constrained by fixed architectures tailored to specific tasks, rigid fusion schemes, and sensitivity to phase misalignment. These limitations motivate the design of a flexible and generalizable enhancement framework that can both selectively preserve and reconstruct speech components while robustly handling phase shifts. This section presents the details of EDNet, our proposed solution.
\subsection{Dual-stream Architecture}
EDNet is designed with a dual-stream architecture to separately process magnitude and phase spectrograms, as shown in ~\Fref{fig:arch}. Instead of using a shared encoder, it employs separate dilated DenseNet encoders~\cite{lu2023explicit} with feature channel dimensions of 64 and 32 for magnitude and phase, respectively. 
This separation is intended to allow magnitude and phase, which have distinct characteristics—magnitude follows a linear scale, where phase exhibits periodicity—to be processed individually, aiming to prevent interference between their estimations.
The decoder also follows this structure, using separate dilated DenseNet decoders~\cite{lu2023explicit} for magnitude and phase. The magnitude decoder directly reconstructs the spectrogram using ReLU activation~\cite{glorot2011deep}, as feature-wise masking is handled within the GM module.

The input speech signal is first transformed using Short-Time Fourier Transform (STFT), decomposing it into magnitude and phase spectrograms. The magnitude spectrogram undergoes power-law compression to enhance feature learning. Each spectrogram is then encoded separately using the respective encoders. Noisy magnitude features are progressively refined through three stacked GM modules. The enhanced magnitude features are then concatenated with the noisy phase features, reduced to 32 channels via a 1×1 convolution, and processed through three stacked Time-Frequency (TF) Mamba blocks~\cite{chao2024investigation}, which model long-range dependencies across temporal and spectral domains using bidirectional Mamba~\cite{gu2024mambalineartimesequencemodeling}, to refine the phase. Finally, the enhanced spectrograms are reconstructed using the decoders, and the enhanced waveform is obtained by combining them with Inverse STFT (ISTFT).

\subsection{Gating Mamba (GM) Module}

The Gating Mamba (GM) module enhances magnitude features by selectively removing non-speech components and reconstructing missing speech information. The GM module consists of multiple GM blocks stacked sequentially. As illustrated in~\Fref{fig:gm_module}, each block receives two inputs: a hidden feature $h_N$, which carries progressively refined speech information from the previous module, and an initial hidden feature $h_0$, which is the original noisy feature from the encoder. In the first block, where no refinement has occurred, both the hidden and initial feature inputs are set to $h_0$, as no previous enhancement step has been applied. This design aims to prevent cumulative degradation during enhancement. As features pass through multiple layers, successive transformations may progressively suppress or distort key speech components. Referencing the original input is intended to mitigate such distortions and better preserve speech integrity. Each block then produces an enhanced feature $h_{N+1}$ based on the input, which is passed to the next module for further refinement.

The module consists of three stages: masking, mapping, and feature fusion. In the masking stage, the hidden feature $h_N$ and initial hidden features $h_0$ are concatenated, passed through a 1×1 convolution for computational efficiency, and processed by a TF-Mamba block. The output is then projected back through a 1×1 convolution, followed by a sigmoid $\sigma$ activation to generate a mask. This mask is multiplied with initial feature $h_0$ to selectively suppress non-speech components while preserving speech-dominant regions. In the mapping stage, the masked feature is concatenated with the hidden feature $h_N$ and processed through another 1×1 convolution followed by a TF-Mamba block. A final 1×1 convolution with a residual connection refines the feature reconstructing the lost speech details. Since masking alone may not fully restore speech—particularly in cases where essential details have been overly suppressed or lost—the mapping process complements it by restoring suppressed or missing speech components. In the feature fusion stage, the outputs from masking and mapping are blended using a weighted sum, where the weights are derived from the masking process. This enables the model to dynamically determine the appropriate balance between suppression and reconstruction, ensuring a progressively refined spectral representation. By iteratively applying this process across GM blocks, the model effectively enhances speech quality by selectively preserving, reconstructing, and refining spectral features.

\subsection{Phase Shift-Invariant Training (PSIT)}

\subsubsection{Phase Shift Estimation}
\label{sec:phase_shift_estimation}
Given a predicted phase spectrogram \(\hat{X}_p(t,f)\) and GT phase \(Y_p(t,f)\), where \( t \) and \( f \) represent the time and frequency indices of the spectrogram, the goal is to determine the optimal shift \(n^*\) that minimizes the phase discrepancy between the predicted phase and the GT. The shifted GT phase is defined as: 
\begin{equation}
    \label{eq:phase_shift}
    Y_p^{\text{shifted}}(t,f,n) = Y_p(t,f) + \frac{2\pi f n}{N},
\end{equation}
where $n$ represents the temporal shift and $N$ is the STFT window length. To estimate the optimal shift, we define the objective function:
\begin{equation}
    E(n) = \sum_{t,f} \left( f_{AW}(\hat{X}_p(t,f) - Y_p^{\text{shifted}}(t,f,n)) \right)^2,
\end{equation}
where $f_{AW}$ is an anti-wrapping function that ensures phase differences remain within a range $[-\pi, \pi]$. However, the presence of the anti-wrapping function introduces non-linearity, making it difficult to derive a closed-form solution for the optimal $n$. 

To address this, we reformulate the problem by applying the anti-wrapping function only to the phase difference between predicted phase and GT, leading to the following objective:
\begin{equation}
    E(n) = \sum_{t,f} \left( f_{AW}(\hat{X}_p(t,f) - Y_p(t,f)) -\frac{2\pi f n}{N} \right)^2.
\end{equation}

This formulation allows us to approximate the optimal shift by assuming a locally linear relationship between the anti-wrapped phase difference and the shift term $2\pi fn/N$. 
For a restricted range of $n$ in $[-0.5, 0.5]$, where shift term remains within the anti-wrapping range, this approximation holds effectively, enabling a differentiable solution. Taking the derivative with respect to $n$ and solving for zero, we obtain:

\begin{equation}
    n^* = \frac{N}{2\pi} \cdot \frac{\sum_{t,f} f \cdot f_{AW}(\hat{X}_p(t,f) - Y_p(t,f))}{\sum_{t,f} f^2}.
\end{equation}
Using this shift, the predicted phase is aligned as:

\begin{equation}
    \hat{X}_p^{\text{aligned}}(t,f) = \hat{X}_p(t,f) - \frac{2\pi f n^*}{N}.
\end{equation}

\subsubsection{Extended Shift Search}
Although the phase shift estimation method provides valid results within the limited range of $[-0.5, 0.5]$, its effectiveness diminishes when the actual phase shift exceeds the range. To address this, we introduce an extended search strategy that evaluates multiple shifts before determining the optimal alignment.

\begin{algorithm}[h]
\footnotesize
\caption{Extended Phase Shift Estimation}
\begin{algorithmic}[1]
\Require $\hat{X}_p$: predicted phase, $Y_p$: GT phase, $\mathcal{S}$: search grid
\Ensure $n^*_{\text{final}}$: optimal shift
\State Initialize $n^*_{\text{final}} \gets 0$, $\text{min\_loss} \gets \infty$
\For {$s \in \mathcal{S}$}
    \State $Y_p^{\text{shifted}} \gets Y_p + \frac{2\pi f s}{N}$
    \State $n^* \gets \frac{N}{2\pi} \cdot \frac{\sum_{t,f} f \cdot f_{AW}(\hat{X}_p - Y_p^{\text{shifted}})}{\sum_{t,f} f^2}$
    \State $\hat{X}_p^{\text{aligned}} \gets \hat{X}_p - \frac{2\pi f n^*}{N}$
    \State $\text{loss} \gets \sum_{t,f} |f_{AW}(\hat{X}_p^{\text{aligned}} - Y_p^{\text{shifted}})|$
    \If {$\text{loss} < \text{min\_loss}$}
        \State $\text{min\_loss} \gets \text{loss}$
        \State $n^*_{\text{final}} \gets n^* + s$
    \EndIf
\EndFor
\State \Return $n^*_{\text{final}}$
\end{algorithmic}
\end{algorithm}
The overall procedure is detailed in Algorithm 1. Instead of directly estimating $n^*$ from the original GT phase $Y_p$, we first produce a set of shifted GT phases using predefined values from a search grid $\mathcal{S}$ ($\{-1, -0.5, 0, 0.5, 1\}$ in this study). For each shifted GT, the optimal phase shift is computed using the phase shift estimation process derived in~\Sref{sec:phase_shift_estimation}, and the predicted phase is realigned accordingly. The alignment quality is then assessed by computing the anti-wrapped L1 distance between the aligned prediction and the corresponding shifted GT phase. The optimal shift $n^{*}$ that minimizes this discrepancy is chosen, and the corresponding optimal phase shift is adjusted accordingly. Note that the predicted phase used in this process is detached from the gradient to prevent unintended backpropagation.

\subsubsection{Loss Computation}
Once the predicted phase $\hat{X}_p$ is aligned using $n^*_{\text{final}}$, any magnitude-phase-based loss function can be directly applied to the aligned phase spectrogram, requiring no modifications beyond the alignment process itself. Following \cite{lu2023explicit}, we adopt the same loss composition, where the total loss is a weighted sum of the PESQ-based GAN discriminator loss, magnitude loss, phase loss, time-domain loss, complex spectrogram loss, and consistency loss. These losses are computed using the predicted magnitude and aligned phase spectrogram, with respective weights of 0.07, 27, 0.3, 6, 3, and 3.

\section{Experiments}
To assess task-specific performance and versatility of the model structure, we conduct experiments on three widely studied single-distortion enhancement tasks: denoising, dereverberation, and bandwidth extension. In addition, we evaluate the model under the more realistic and challenging multi-distortion conditions, where multiple distortions are combined. 

\subsection{Datasets} 
\subsubsection{Denoising}
To assess speech denoising performance, we adopt the VoiceBank+DEMAND dataset~\cite{valentini2016investigating}, a standard benchmark widely used in prior research. The clean speech samples originate from the VoiceBank corpus~\cite{veaux2017vctk}, comprising 11,572 utterances from 28 speakers for training, and 824 utterances from 2 unseen speakers for testing. Noise is synthetically added using 10 distinct noise types—8 sourced from the DEMAND database~\cite{thiemann2013diverse} and 2 artificially generated—at signal-to-noise ratios (SNRs) of 0, 5, 10, and 15 dB. The test set contains 5 previously unseen noise conditions at SNRs of 2.5, 7.5, 12.5, and 17.5 dB. All audio data is uniformly resampled to a 16 kHz sampling rate.

\subsubsection{Dereverberation}
For evaluating dereverberation capabilities, we utilize the dataset provided by the REVERB Challenge~\cite{kinoshita2016summary}. The training portion comprises 7,861 reverberant utterances, generated by convolving clean WSJCAM0~\cite{fransen1994wsjcam0} recordings with real room impulse responses (RIRs) and embedding ambient noise at an SNR of 20 dB. Reverberation scenarios cover three room sizes with 60 dB reverberation time (RT60) values of 0.3, 0.6, and 0.7 seconds, each captured at near (0.5 m) and far (2.0 m) microphone placements. The evaluation set consists of 2,176 simulated utterances and 372 real recordings, sourced from the MC-WSJ-AV corpus~\cite{lincoln2005multi}. For consistency, we use the single-channel version of all recordings at 16 kHz.

\subsubsection{Bandwidth Extension}
For the bandwidth extension task, we utilize the VCTK corpus~\cite{veaux2017vctk}, which comprises approximately 44 hours of high-quality speech data from 108 English speakers, originally recorded at a 48 kHz sampling rate. Our data preparation procedure follows established protocols from prior work~\cite{kuleshov2017audio}. Specifically, we allocate the first 100 speakers for training and reserve the remaining 8 for evaluation. To construct input-target pairs, we begin by downsampling the original 48 kHz recordings to 16 kHz, designating these as wideband reference signals. Narrowband inputs are then simulated by further subsampling the audio with factors of 2 and 4, limiting the bandwidth to 8 kHz and 4 kHz, respectively. These narrowband signals are subsequently upsampled back to 16 kHz using spline interpolation, serving as the model input. The corresponding wideband audio remains the ground truth target for supervision.

\subsubsection{Multi-distortion Enhancement}
Since no standard benchmark exists for multi-distortion enhancement, we synthesize the data by introducing three types of distortions: additive noise, reverberation, and bandwidth limitation. 
Noise is mixed using the DEMAND database~\cite{thiemann2013diverse}, with a randomly selected SNR ranging between -6 and 14 dB, following the same noise-type split as the VoiceBank+DEMAND dataset~\cite{valentini2016investigating}. Reverberation is applied using room impulse responses generated by Pyroomacoustics, with room dimensions varying between 5 and 15 m in length and width, and 2 to 6 m in height. The RT60 is randomly selected between 0.4 and 1.0 seconds. Bandwidth limitation is applied using randomly selected low-pass filters, including Butterworth, Bessel, and Chebyshev filters, with cutoff frequencies set to 2, 4, or 8 kHz. We use the VCTK dataset~\cite{veaux2017vctk} as a speech source, selecting the first 100 speakers for training and the last 8 for testing. Speakers p280 and p310 are excluded due to technical issues.

\subsection{Evaluation Metrics}
To assess the effectiveness of the proposed model across various tasks, we employ a range of standard evaluation metrics commonly used in speech enhancement. Especially for single-distortion tasks, we follow the benchmarking conventions established in prior work for each dataset to facilitate direct comparison with existing methods.

In the denoising task, we use the perceptual evaluation of speech quality (PESQ), short-time objective intelligibility (STOI), and three composite metrics: CSIG, CBAK, and COVL~\cite{hu2007evaluation}. PESQ provides an estimate of overall perceptual quality, while STOI quantifies intelligibility. The composite scores CSIG, CBAK, and COVL assess speech distortion, background noise intrusiveness, and overall perceptual quality, respectively. 

In the dereverberation task, we adopt the official evaluation toolkit provided by the REVERB Challenge~\cite{kinoshita2016summary}, which measuring various metrics including PESQ, cepstral distance (CD), log-likelihood ratio (LLR), and frequency-weighted segmental SNR (FWSegSNR). These metrics capture perceptual speech quality, time-frequency domain distortions, spectral divergence, and residual reverberation energy. Since real-recorded reverberant utterances do not have clean references, the signal-to-reverberation modulation energy ratio (SRMR) is used exclusively for those cases, serving as a non-intrusive proxy for dereverberation effectiveness.

For the bandwidth extension task, we utilize PESQ to assess the perceptual restoration of missing high-frequency components, ViSQOL (Virtual Speech Quality Objective Listener)~\cite{chinen2020visqol} to reflect MOS-LQO (mean opinion score - listening quality objective), and log-spectral distance (LSD)~\cite{gray1976distance} to quantify spectral distortion. 

In the multi-distortion task, where multiple degradation types (noise, reverberation, and bandwidth limitation) co-occur, a broader and more comprehensive set of metrics is adopted. We report PESQ, CSIG, CBAK, COVL, and STOI to assess perceptual and intelligibility performance~\cite{hu2007evaluation}. Additionally, ViSQOL~\cite{chinen2020visqol} is used to gauge overall listening quality, while LSD~\cite{gray1976distance} complements the evaluation by quantifying spectral accuracy.  Finally, perceptual speaker similarity (PSS) is assessed by computing the cosine similarity between speaker embeddings of the enhanced and clean utterances. These embeddings are extracted using a WavLM-ECAPA model trained on the VoxSim dataset~\cite{ahn2024voxsim}, a subset of the VoxCeleb1 dataset~\cite{nagrani2020voxceleb} annotated with human-labeled perceptual similarity scores.

\subsection{Implementation Details}
All speech signals are sampled at 16 kHz. During training, random 2-second segments are extracted from each utterance. STFT is computed using a 400-point FFT, a 25 ms window (400 samples), and a 6.25 ms hop size (100 samples). Magnitude spectrograms are min-max normalized before input. All models are trained for 200 epochs using AdamW optimizer~\cite{loshchilov2017decoupled}, with a batch size of 8 and learning rate of 0.0007, decayed exponentially by 0.99 per epoch. We track validation PESQ every 2,000 steps and select the best checkpoint. For a fair comparison and a focused analysis of the model architecture and learning objectives, we exclude specialized techniques from all experiments such as Perceptual Contrast Stretching (PCS)~\cite{chao2022perceptual} that can boost objective score.

For phase reconstruction, we modify the model to isolate the phase stream. Specifically, the GM module is removed, and clean magnitude features from the magnitude encoder are directly fed into the phase stream. The feature size of the phase stream is set to 64. To enable statistically meaningful analysis, each PSIT configuration is trained for 80 epochs, 15 times with different random seeds, under identical settings.

\section{Experimental Results and Analysis}

\subsection{Single-Distortion Benchmark Comparison}
\begin{table}[t]
    \centering
    \caption{Comparison of Baselines for Denoising.}
    \vspace{-2mm}
    \resizebox{0.9\linewidth}{!}{
    \begin{tabular}{lccccc}
        \toprule
        \textbf{Method}  & \textbf{PESQ$\uparrow$} & \textbf{CSIG$\uparrow$} & \textbf{CBAK$\uparrow$} & \textbf{COVL$\uparrow$} & \textbf{STOI$\uparrow$} \\
        \midrule
        Noisy   & 1.97 & 3.35 & 2.44 & 2.63 & 0.91  \\
        \midrule
        DEMUCS \cite{defossez2020real}   & 3.07 & 4.31 & 3.40 & 3.63 & 0.95  \\
        CMGAN \cite{abdulatif2024cmgan}  & 3.41 & 4.63 & 3.94 & 4.12 & \textbf{0.96}  \\
        TridentSE \cite{yin2022tridentse}  & 3.47 & 4.70 & 3.81 & 4.10 & \textbf{0.96}  \\
        SEMamba \cite{chao2024investigation}   & 3.55 & 4.77 & 3.95 & 4.26 & \textbf{0.96} \\
        MP-SENet \cite{lu2023explicit}   & \textbf{3.60} & \textbf{4.81} & \textbf{3.99} & \textbf{4.34} & \textbf{0.96} \\
        \midrule
        \textbf{EDNet} w/o PSIT   & 3.55 & 4.75 & 3.83 & 4.28 & \textbf{0.96} \\
        \textbf{EDNet}    & 3.58 & 4.78 & 3.85 & 4.31 & \textbf{0.96} \\
        \bottomrule
    \end{tabular}
    }
    \label{tab:denoise}
    \vspace{-3mm}
\end{table}
\begin{table}[t]
    \centering
    \caption{Comparison of Baselines for Dereverberation.}
    \vspace{-2mm}
    \resizebox{0.99\linewidth}{!}{
    \begin{tabular}{lcccccc}
        \toprule
        \multirow{2}{*}{\textbf{Method}}  & \multicolumn{5}{c}{\textbf{SimData}} & \textbf{RealData} \\
        \cmidrule(lr){2-6}  \cmidrule(lr){7-7}
         & PESQ$\uparrow$ & CD$\downarrow$ & LLR$\downarrow$ & SNR$_{fw}$$\uparrow$ & SRMR$\uparrow$& SRMR$\uparrow$  \\
        \midrule
        Reverberant   & 1.50 & 3.97 & 0.58 & 3.62 & 3.69 & 3.18 \\
        \midrule
        WPE \cite{nakatani2010speech}   & 1.72 & 3.75 & 0.51 & 4.90 & 4.22 & 3.98 \\
        UNet \cite{ernst2018speech}  & - & 2.50 & 0.40 & 10.70 & 4.88 & 5.58 \\
        CMGAN \cite{abdulatif2024cmgan}  & - & 2.25 & 0.31 & 11.74 & 5.47 & 6.55 \\
        DCN \cite{kothapally2024monaural}   & 2.94 & 2.00 & \textbf{0.23} & 13.33 & 5.27 & 6.48\\
        MP-SENet \cite{lu2023explicit}   & 2.97 & 1.97 & 0.24 & 14.07 & 5.51 & 6.67\\
        \midrule
        \textbf{EDNet} w/o PSIT  & 3.06 & 1.95 & \textbf{0.23} & 13.78 & 5.59 & \textbf{7.42}\\
        \textbf{EDNet}  & \textbf{3.17} & \textbf{1.87} & \textbf{0.23} & \textbf{14.14} & \textbf{5.63} & 7.22\\
        \bottomrule
    \end{tabular}
    }
    \label{tab:dereverb}
    \vspace{-3mm}
\end{table}
\begin{table}[t]
    \centering
    \caption{Comparison of Baselines for Bandwidth Extension.}
    \vspace{-2mm}
    \resizebox{0.99\linewidth}{!}{
    \begin{tabular}{lccc|cccc}
        \toprule
        \multirow{2}{*}{\textbf{Method}}  & \multicolumn{3}{c|}{\textbf{8 kHz → 16 kHz}} & \multicolumn{3}{c}{\textbf{4 kHz → 16 kHz}}\\
        \cmidrule(lr){2-4} \cmidrule(lr){5-7}
        & PESQ$\uparrow$ & LSD$\downarrow$ & ViSQOL$\uparrow$ & PESQ$\uparrow$ & LSD$\downarrow$ & ViSQOL$\uparrow$ \\
        \midrule
        AFiLM \cite{rakotonirina2021self}  & - & 1.24 & 4.39 & - & 1.63 & 3.83 \\
        AECNN \cite{wang2021towards}  & 3.91 & 0.88 & - & 3.64 & 0.95 & - \\
        NVSR \cite{liu2022neural}  & 3.56 & 0.79 & 4.52 & 2.40 & 0.95 & 4.11 \\
        AP-BWE \cite{lu2024towards} & - & 0.69 & 4.71 & - & 0.87 & 4.30 \\
        MP-SENet \cite{lu2023explicit}  & 4.28 & 0.66 & 4.72 & 3.78 & 0.81 & 4.42 \\
        \midrule
        \textbf{EDNet} w/o PSIT & 4.35 & 0.64 & 4.75 & 3.87 & 0.79 & 4.42\\
        \textbf{EDNet}  & \textbf{4.37} & \textbf{0.62} & \textbf{4.77} & \textbf{3.88} & \textbf{0.78} & \textbf{4.43}\\
        \bottomrule
    \end{tabular}
    }
    \label{tab:bandext}
    \vspace{-4mm}
\end{table}
\newpara{Denoising:}~\Tref{tab:denoise} presents EDNet’s performance on the denoising task. EDNet achieves competitive results compared to state-of-the-art models, demonstrating strong performance across key metrics.
Without PSIT, its performance is comparable to that of SEMamba~\cite{chao2024investigation}, which shares a similar architectural foundation, combining a DenseNet-based encoder-decoder with a TF-Mamba backbone. However, it still underperforms MP-SENet~\cite{lu2023explicit}, as the GM module applies masking at the feature level rather than directly in the STFT domain, which likely limits its ability to preserve fine-grained signal details. The inclusion of PSIT yields consistent improvements in most of the key metrics. These gains suggest that PSIT can contribute to performance even in denoising scenarios, where temporal misalignment and phase distortion are relatively limited compared to other distortion types. Although STOI scores are saturated across top-performing models (0.96), the observed improvements in perceptual metrics indicate that PSIT helps produce more natural and intelligible outputs. Overall, EDNet demonstrates solid performance in denoising and serves as a strong baseline, though it has not yet surpassed all existing models, leaving room for further improvement.

\newpara{Dereverberation:}~\Tref{tab:dereverb} shows that EDNet achieves superior performance on both simulated and real reverberant speech. On the simulated set, it achieves the best results across all key metrics, including the highest PESQ, the lowest CD and LLR, along with substantial improvements in FWSegSNR and SRMR. On real recordings, EDNet also achieves the highest SRMR among all prior models, although its variant without PSIT records a slightly higher score. Though the inclusion of PSIT results in a minor SRMR drop, it consistently yields substantial improvements in other perceptual metrics, including PESQ, CD, and FWSegSNR. These results suggest that PSIT is particularly beneficial in reverberant conditions, where phase distortion and temporal misalignment are especially pronounced due to the nature of reverberation.

\newpara{Bandwidth extension:} As shown in~\Tref{tab:bandext}, EDNet outperforms prior work in both 8 kHz → 16 kHz and 4 kHz → 16 kHz bandwidth extension tasks. It achieves the highest PESQ scores and the lowest LSD, indicating accurate spectral reconstruction. ViSQOL scores also surpass those of prior work by a small margin, reinforcing the perceptual quality gains achieved by EDNet. 
While the gains from PSIT are not large, they are consistent across all metrics, suggesting potential benefits in bandwidth extension scenarios, where temporal misalignment is limited but phase information in the high-frequency range is often absent. Given the nature of this distortion, PSIT may offer a more appropriate supervision under such conditions, as high-frequency phase components are inherently more sensitive to temporal shifts.

In summary, EDNet achieves state-of-the-art or competitive results across denoising, dereverberation, and bandwidth extension benchmarks. These results are obtained without task-specific modifications or hyperparameter tuning, demonstrating strong structural flexibility and adaptability. This adaptability is most pronounced in the dereverberation task. While mapping-based reconstruction is commonly favored for dereverberation given its convolutive distortion model~\cite{lemercier2023extending, kothapally2024monaural}, a single method may not be uniformly optimal across all conditions. Depending on the reverberation level, speech content may remain partially intact, favoring suppression-based strategies, whereas severe smearing can necessitate reconstruction. The strength of EDNet lies in the architectural capacity to adaptively balance these strategies rather than rely on a fixed processing paradigm. Strong results against competitive baselines from both methodological families, including the mapping-based DCN and the masking-oriented MP-SENet, support the effectiveness of this data-driven hybrid design. Beyond architectural design, PSIT consistently improves performance across various distortion types. The most substantial gains are observed in dereverberation, where temporal misalignment is pronounced. It also provides incremental but consistent benefits in scenarios involving different phase-related degradations, such as additive noise and bandwidth limitation.

\begin{table*}[!t]
    \centering
    \caption{Comparison of Different Methods for Multi-distortion Enhancement.}
    \vspace{-2mm}
    \resizebox{0.75\linewidth}{!}{
    \begin{tabular}{lcccccccccc}
        \toprule
        \textbf{Method} & \textbf{Year} & \textbf{Param.} & \textbf{PESQ$\uparrow$} & \textbf{CSIG$\uparrow$} & \textbf{CBAK$\uparrow$} & \textbf{COVL$\uparrow$} & \textbf{STOI$\uparrow$} & \textbf{LSD$\downarrow$} & \textbf{ViSQOL$\uparrow$} & \textbf{PSS$\uparrow$} \\
        \midrule
        Noisy  & - & -& 1.28 & 1.16 & 1.82 & 1.16 & 0.51 &  3.64 & 2.40 & 0.373 \\
        \midrule
        VoiceFixer\cite{liu2022voicefixer} & 2022 & 122M & 1.39 & 2.24 & 2.00 & 1.86 & 0.61 & 2.48 & 3.11  & 0.604  \\
        UNIVERSE\cite{serra2022universal} & 2022 & 107M & 1.38 & 2.11 & 2.16 & 1.78 & 0.54 & 2.73 & 2.94  & 0.611 \\
        HD-DEMUCS\cite{kim2023hd} & 2023 & 23.6M & 1.38 & 2.69 & 2.20 & 2.08 & 0.64 & 2.22 & 3.28 & 0.575 \\
        UNIVERSE++\cite{scheibler2024universal} & 2024 & 107M & 1.50 & 2.24 & 2.27 &  1.91 & 0.61 & 2.74 & 3.13 & 0.678 \\
        CMGAN\cite{abdulatif2024cmgan} & 2024 & 1.83M & 1.80 & 2.20 & 2.47 & 2.04 & 0.71 & 2.40 & 3.30 & 0.552 \\
        SEMamba \cite{chao2024investigation}   & 2024 & 2.25M & 2.26 & 2.46 & 2.79 & 2.41 & 0.73 &  2.38 &  3.33 & 0.596  \\
        MP-SENet \cite{lu2023explicit}  & 2025 &2.26M & 2.24 & 2.63 & 2.75 & 2.49 & 0.72 & 2.31 &  3.25 & 0.604\\
        \midrule
        \textbf{EDNet} w/o PSIT & 2025 & 2.67M & 2.32 & 3.02 & 2.77 & 2.72 & 0.74 & 2.19 &  3.52 & 0.667 \\
        \textbf{EDNet} & 2025 & 2.67M & \textbf{2.39} & \textbf{3.06} & \textbf{2.81} & \textbf{2.78} & \textbf{0.75} & \textbf{2.16} & \textbf{3.60} & \textbf{0.693}  \\
        \bottomrule
    \end{tabular}
    }
    \label{tab:3dist}
    \vspace{-2mm}
\end{table*}

\begin{table}[!t]
    \centering
    \caption{Results of Subjective Listening Test for Multi-distortion Enhancement. SQ: Speech Quality, CP: Content Preservation}
    \vspace{-2mm}
    \resizebox{0.99\linewidth}{!}{
    \begin{tabular}{l|c|c|cc|cc}
        \toprule
        \textbf{Method} & \textbf{Type} & \textbf{PESQ$\uparrow$} & \textbf{DNSMOS$\uparrow$} & \textbf{NISQA$\uparrow$} & \textbf{SQ$\uparrow$} & \textbf{CP$\uparrow$} \\
        \midrule
        VoiceFixer\cite{liu2022voicefixer} & G & 1.39 & 2.897 & 3.047 & 2.26 $\pm$ 0.09 & 2.36 $\pm$ 0.07 \\
        UNIVERSE\cite{serra2022universal} & G & 1.38 & 2.821 & 4.081 & 2.40 $\pm$ 0.09 & 1.83 $\pm$ 0.04\\
        UNIVERSE++\cite{scheibler2024universal} & G & 1.50 & 2.897 & \textbf{4.216} & 2.70 $\pm$ 0.08 & 2.29 $\pm$ 0.06 \\
        \midrule
        HD-DEMUCS\cite{kim2023hd} & D & 1.38 & 2.712 & 1.240 & 1.41 $\pm$ 0.05 & 2.04 $\pm$ 0.05\\
        CMGAN\cite{abdulatif2024cmgan} & D & 1.80 & 2.968 & 2.507 & 2.07 $\pm$ 0.08 & 3.00 $\pm$ 0.06 \\
        SEMamba \cite{chao2024investigation}   & D & 2.26 & 2.968 & 2.885 & 2.56 $\pm$ 0.07 & 3.53 $\pm$ 0.06 \\
        MP-SENet \cite{lu2023explicit}  & D & 2.24 & 2.953 & 2.804 & 2.58 $\pm$ 0.08 & 3.53 $\pm$ 0.06 \\
        \midrule
        \textbf{EDNet} & D & \textbf{2.39} & \textbf{2.981} & 2.967 & \textbf{2.73 $\pm$ 0.07} & \textbf{3.58 $\pm$ 0.06} \\
        \bottomrule
    \end{tabular}
    }
    \label{tab:mos}
    \vspace{-2mm}
\end{table}
\subsection{Multi-Distortion Performance Comparison}
For multi-distortion performance comparison, we first consider representative multi-distortion speech enhancement models that explicitly aim to handle diverse distortion types—VoiceFixer \cite{liu2022voicefixer}, UNIVERSE~\cite{serra2022universal}, HD-DEMUCS~\cite{kim2023hd}, and UNIVERSE++ \cite{scheibler2024universal}. In addition to these multi-distortion enhancement baselines, we also include CMGAN~\cite{abdulatif2024cmgan}, SEMamba~\cite{chao2024investigation}, and MP-SENet~\cite{lu2023explicit}, which are architecturally and functionally similar to EDNet. These time–frequency domain models employ dilated DenseNet-based encoder–decoders with temporal–frequency modeling backbones such as TF-Conformer, Transformer, or Mamba. While not all target multi-distortion enhancement explicitly, their unified treatment of diverse distortions makes them suitable for evaluating generalization. All baselines offer publicly available implementations, ensuring reproducible benchmarking under consistent settings. Since VoiceFixer is designed for 44.1 kHz full-band audio, retraining it at 16 kHz is not straightforward due to architectural and preprocessing constraints. Therefore, we use its official checkpoint for inference and downsample the generated 44.1 kHz outputs to 16 kHz for evaluation. Although this setup may not represent a perfectly aligned condition, it still provides a meaningful reference for comparison, as the released checkpoint is trained with identical speech and noise sources.

As summarized in~\Tref{tab:3dist}, EDNet achieves the best performance across every evaluation metric. While detailed analyses are presented in following~\Sref{sec:gm_ablation},~\ref{sec:gate_analysis} and~\ref{sec:psit}, the performance gains appear to be related to the complementary roles of the GM module and PSIT. The GM module employs a gating mechanism to dynamically blend masking and mapping strategies in a region-wise manner, allowing the network to emphasize the most beneficial features based on input characteristics. PSIT also contributes to performance gains in the multi-distortion setting, where diverse sources of phase degradation—such as misalignment, noise-induced perturbations, and loss of high-frequency phase information—coexist. The shift-tolerant nature of PSIT may be better aligned with these conditions, potentially allowing it to provide more consistent guidance during training. Notably, even in the absence of PSIT, EDNet consistently outperforms existing baselines across key metrics, highlighting the strength of its architectural design in multi-distortion settings.

In addition to objective evaluations, we also conduct subjective listening tests to complement the quantitative results and to better understand the perceptual differences that may not be captured by conventional metrics. Recent studies~\cite{liu2022neural, zhang2024urgent, richter2023speech} report that generative models often produce perceptually high-quality outputs despite achieving lower scores on reference-based objective metrics such as PESQ. To further analyze this discrepancy, we design two complementary listening tests: (1) speech quality test, where participants rate the perceptual quality of enhanced signals without reference access, and (2) content preservation test, where they compare the enhanced and ground-truth speechs to assess the degree of content consistency, including speaker similarity and linguistic alignment. Each test involve 20 participants and 15 randomly selected utterances from the test set. To further align these subjective assessments with objective measures, we report DNSMOS~\cite{reddy2021dnsmos} and NISQA~\cite{mittag2021nisqa} as reference-free metrics.

As summarized in~\Tref{tab:mos}, generative models generally receive speech quality scores that are higher than expected from their reference-based objective metrics, highlighting a clear discrepancy between perceptual and quantitative evaluations. This suggests that their outputs often sound more natural and artifact-free than the reference-based objective scores alone indicate, a trend confirmed by their superior NISQA scores compared to all discriminative baselines. However, the same models—particularly diffusion-based systems such as UNIVERSE and UNIVERSE++—exhibit substantially lower content preservation scores than discriminative baselines, indicating frequent modifications of linguistic content or speaker identity, which likely explains their relatively low reference-based metric results. By contrast, discriminative models preserve content more faithfully but are observed to introduce high-frequency artifacts when restoring severely degraded regions, which is reflected in their relatively lower NISQA performance. Taken together, these results highlight an inherent trade-off between perceptual naturalness and content fidelity in current speech enhancement paradigms and underscore the limitations of relying solely on objective metrics for direct comparisons between discriminative and generative approaches.

EDNet achieves the best scores on both speech quality and content preservation, indicating high perceptual quality with strong content fidelity. In terms of reference-free metrics, it achieves the highest DNSMOS score overall and the highest NISQA score among the discriminative models. While its NISQA performance is still lower than that of generative models—likely due to high-frequency artifacts that are not fully resolved—EDNet shows a clear improvement over other discriminative baselines. These results suggest that our architectural design, which combines dynamic gated fusion with phase-tolerant supervision, successfully enhances perceptual quality while maintaining the reliable fidelity inherent in discriminative methods. Taken together, these results suggest a practical path to unifying the perceptual strengths of generative models with the content fidelity of discriminative methods. We further provide audio samples generated by our method on the demo page\footnote{\href{https://mm.kaist.ac.kr/projects/EDNet}{https://mm.kaist.ac.kr/projects/EDNet}}.

\subsection{GM Module Ablation Study}
\label{sec:gm_ablation}
\begin{table*}[t]
    \centering
    \caption{Results of GM Module Ablation Study.}
    \vspace{-2mm}
    \resizebox{0.7\linewidth}{!}{
    \begin{tabular}{lcccccccccc}
        \toprule
        \textbf{Method} & \textbf{PESQ$\uparrow$} & \textbf{CSIG$\uparrow$} & \textbf{CBAK$\uparrow$} & \textbf{COVL$\uparrow$} & \textbf{STOI$\uparrow$} & \textbf{LSD$\downarrow$} & \textbf{ViSQOL$\uparrow$} & \textbf{PSS$\uparrow$}\\
        \midrule
        \textbf{EDNet} & \textbf{2.39} & \textbf{3.06} & \textbf{2.81} & \textbf{2.78} & \textbf{0.75} & \textbf{2.16} & \textbf{3.60} & \textbf{0.693}\\
        \midrule
         (\textit{A}) w/o masking module & 2.27 & 2.96 & 2.74 & 2.67 & 0.74 & 2.19 &  3.53 & 0.667 \\
         (\textit{B}) w/o mapping module & 2.11 & 2.81 & 2.62 & 2.52 & 0.71 & 2.22 &  3.37 & 0.604 \\
         (\textit{C}) w/o feature fusion module & 2.24 & 3.02 & 2.70 & 2.68 & 0.72 & \textbf{2.16} &  3.41 & 0.638 \\
         (\textit{D}) w/o initial feature $h_0$ & 2.35 & 2.93 & 2.79 & 2.69 & \textbf{0.75} & 2.19 &  3.59 & 0.686  \\
         (\textit{E}) w/o masked feature condition & 2.33 & 2.96 & 2.78 & 2.70 & 0.74 & 2.18 &  3.54 & 0.689 \\
         (\textit{F}) w/ fusion weight from mapping module & 2.25 & 2.95 & 2.70 & 2.65 & 0.72 & 2.21 &  3.34 & 0.643 \\
        \bottomrule
    \end{tabular}
    }
    \label{tab:gm_ablation}
    
    \vspace{-4mm}
\end{table*}
To validate the design choices of the GM module, we perform an ablation study by systematically disabling or modifying its internal components. The results are summarized in~\Tref{tab:gm_ablation}. 
We first evaluate the impact of removing either the masking or mapping branch. Removing the masking branch results in moderate performance degradation (row \textit{A}), while eliminating the mapping branch leads to a substantial drop across all metrics (row \textit{B}). This highlights the importance of the mapping branch in generative tasks, such as bandwidth extension, where reconstructing missing spectral content is essential. In contrast, the masking branch helps preserve fine-grained details and performs well in scenarios where most of the original signal remains intact but is locally degraded, such as in denoising and dereverberation. However, under more challenging conditions involving simultaneous noise, reverberation, and bandwidth limitation, masking alone proves inadequate, emphasizing the necessity of a generative component.
Next, we evaluate whether a simple combination of the two branches without dynamic gating can effectively leverage their strengths (row \textit{C}). When masking and mapping outputs are averaged using a fixed fusion weight of 0.5, performance improves over using masking alone but falls short of the mapping-only baseline. This suggests that heuristic fusion mechanism can result in an averaging effect, rather than a meaningful integration of complementary features.

To further explore the design space, we evaluate three additional variations that retain the gating mechanism but alter the flow of information. 
First, to investigate the role of the initial input feature $h_0$, we remove its connection to the GM module so that each layer relies solely on the hidden representation of the previous layer (row \textit{D}). This modification leads to reduced final performance and increased instability during early training, with some runs failing to converge entirely. The performance reported in the table corresponds only to successful training runs. A plausible explanation is that representational degradation accumulates across layers; immature gating and masking in the early training phase may cause excessive information loss and hinder gradient propagation. Maintaining a connection to the initial feature appears to stabilize training in the early stages,
which ultimately leads to improved final performance. Second, we decouple the two branches by removing the conditioning connection from the masking to the mapping module (row \textit{E}). This independent-path configuration also reduces performance, indicating that cross-branch interaction provides meaningful cues for enhancing reconstruction. Finally, we modify the fusion weight to be predicted by the mapping module instead of the masking module (row \textit{F}). This configuration yields even lower performance than fixed average gating. One possible explanation is that the gating function is structurally and functionally more aligned with the masking branch, making it less compatible with the objectives of the mapping branch. This mismatch may degrade both the effectiveness of the gating and the performance of the mapping process.

Taken together, these results show that the GM module is not merely an ensemble of two strategies, but a tightly integrated design. Optimal performance depends critically on how fusion is performed, where it is positioned, and how the branches interact through shared and conditional information.

\subsection{Gate Behavior Analysis}
\label{sec:gate_analysis}

To examine how the Gating Mamba (GM) module dynamically allocates enhancement effort, we analyze its gating behavior across different tasks and input conditions. 
We compute the average gate activation across feature channels and interpolate it to match the resolution of the input spectrogram. 
These visualization provide insights into the spatial distribution of masking and mapping emphasis within the model.

\subsubsection{Task-Specific Behavior}
\begin{table}[t]
    \centering
    \caption{Mean Gate Values for Task-specific Scenarios.}
    \vspace{-2mm}
    \resizebox{0.8\linewidth}{!}{
    \begin{tabular}{llc}
        \toprule
        \textbf{Task} & \textbf{Test Dataset} & \textbf{Gate Mean} \\
        \midrule
         Denoising & VoiceBank+DEMAND & 0.772  \\
         Dereverberation & REVERB & 0.693 \\
         Bandwidth Extension & VCTK 8 kHz & 0.760 \\
         Bandwidth Extension   & VCTK 4 kHz & 0.731 \\
        \bottomrule
    \end{tabular}
    }
    \label{tab:gate_mean_task}
    \vspace{-4mm}
\end{table}
\begin{figure*}[!t]
    \centering
    \subfloat[\rmfamily\footnotesize Source Speech]{
        \includegraphics[width=0.212\textwidth]{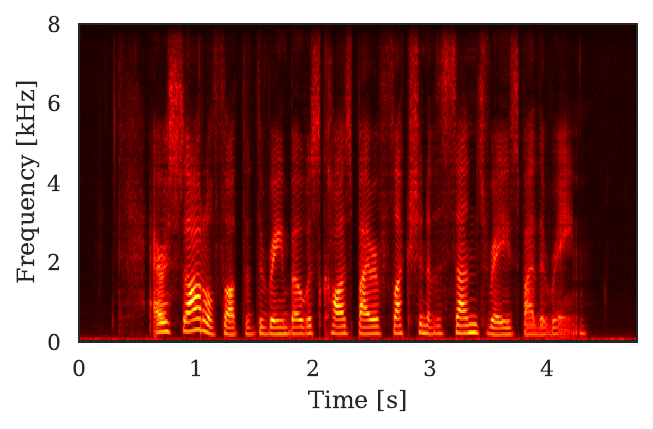}
        \label{fig:gvis_task_a}
    }
    \hfill
    \subfloat[\rmfamily\footnotesize Additive Noise]{
        \includegraphics[width=0.23\textwidth]{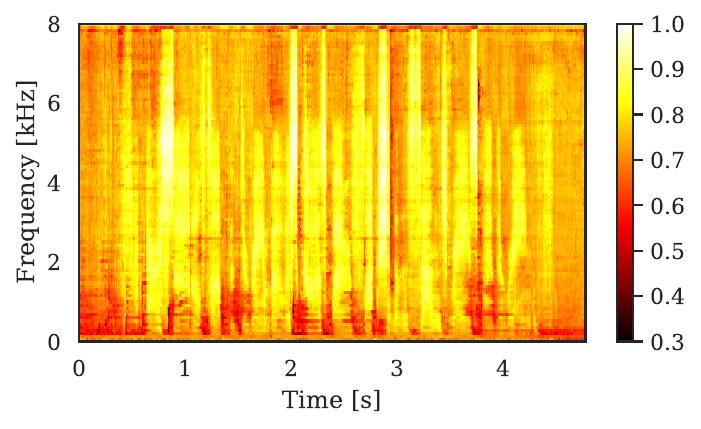}
        \label{fig:gvis_task_b}
    }
    \hfill
    \subfloat[\rmfamily\footnotesize Reverberation]{
        \includegraphics[width=0.23\textwidth]{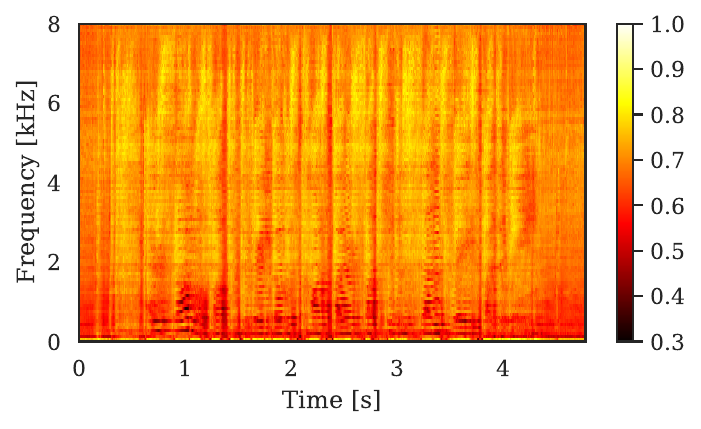}
        \label{fig:gvis_task_c}
    }
    \hfill
    \subfloat[\rmfamily\footnotesize Bandwidth Limitation 8 kHz]{
        \includegraphics[width=0.23\textwidth]{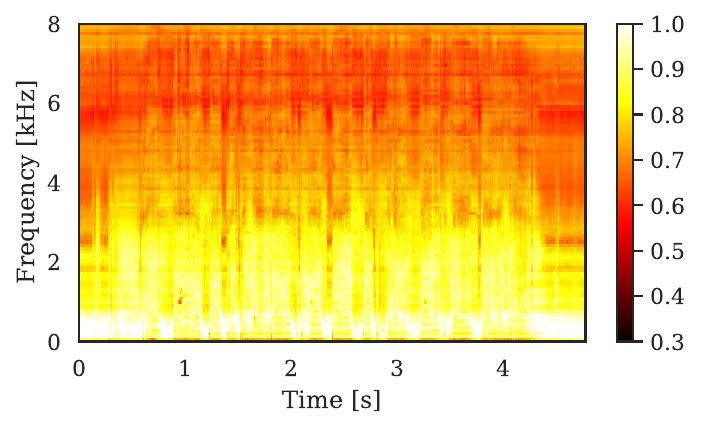}
        \label{fig:gvis_task_d}
    }

    \caption{Visualization of gate maps in task-specific scenarios using single-distortion models. (a) Clean source speech. Gate maps of (b) the denoising model given additive noise input, (c) the dereverberation model given reverberant input, and (d) the bandwidth extension model given 8 kHz bandwidth-limited input.}
    \vspace{-5mm}
    \label{fig:gvis_task}
\end{figure*}

\begin{table}[t]
    \centering
    \caption{Mean Gate Values for Input-specific Scenarios.}
    \vspace{-2mm}
    \resizebox{0.55\linewidth}{!}{
    \begin{tabular}{lc}
        \toprule
        \textbf{Test Dataset} & \textbf{Gate Mean} \\
        \midrule
         Full Dataset  & 0.582  \\
         \midrule
         \hspace{0.2em}- cut-off frequency 8 kHz & 0.594 \\
         \hspace{0.2em}- cut-off frequency 4 kHz & 0.583 \\
         \hspace{0.2em}- cut-off frequency 2 kHz & 0.569 \\
         \hspace{0.2em}- SNR=(-6,0) & 0.582 \\
         \hspace{0.2em}- SNR=(0,6) & 0.582 \\
         \hspace{0.2em}- SNR=(6,12) & 0.583 \\
        \bottomrule
    \end{tabular}
    }
    \label{tab:gate_mean_input}
    \vspace{-4mm}
\end{table}
\begin{figure*}[!t]
   \centering
    \centering
    \subfloat[\rmfamily\footnotesize Additive Noise]{
       \begin{minipage}{0.23\textwidth}
            \centering
            \includegraphics[width=\linewidth]{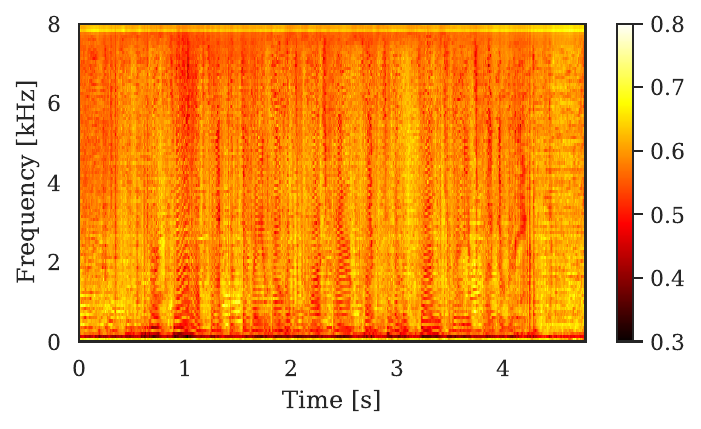}\\
            \includegraphics[width=0.9\linewidth]{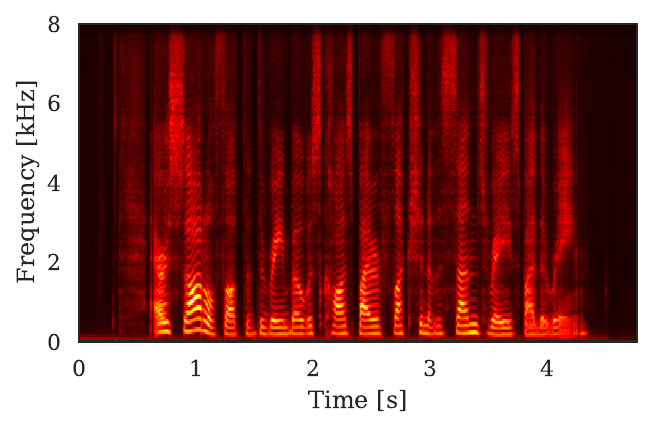}
            \label{fig:gvis_input_a}
            \end{minipage}
    }
    \hfill
    \subfloat[\rmfamily\footnotesize Reverberation]{
       \begin{minipage}{0.23\textwidth}
            \centering
            \includegraphics[width=\linewidth]{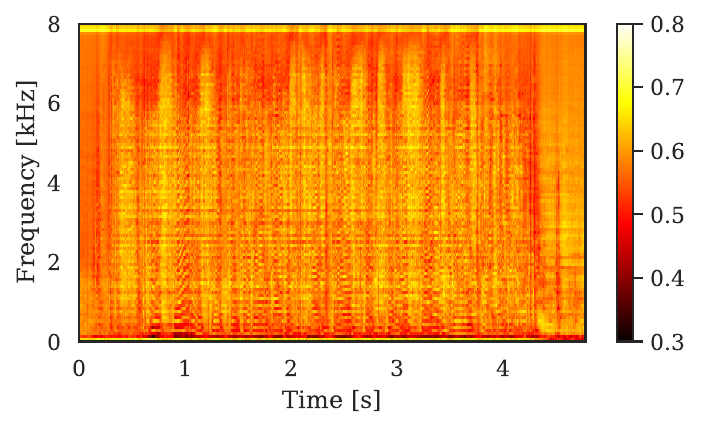}\\
            \includegraphics[width=0.9\linewidth]{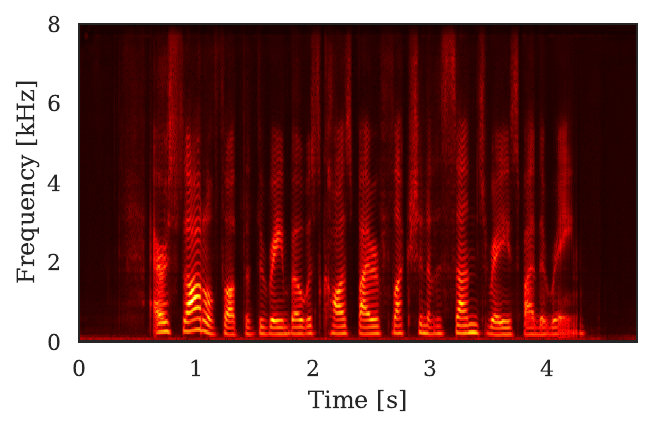}
            \label{fig:gvis_input_b}
            \end{minipage}
    }
    \hfill
    \subfloat[\rmfamily\footnotesize Bandwidth Limitation 8 kHz]{
       \begin{minipage}{0.23\textwidth}
            \centering
            \includegraphics[width=\linewidth]{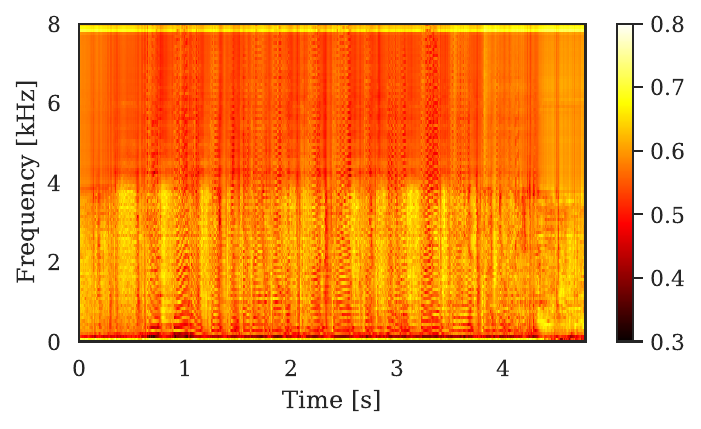}\\
            \includegraphics[width=0.9\linewidth]{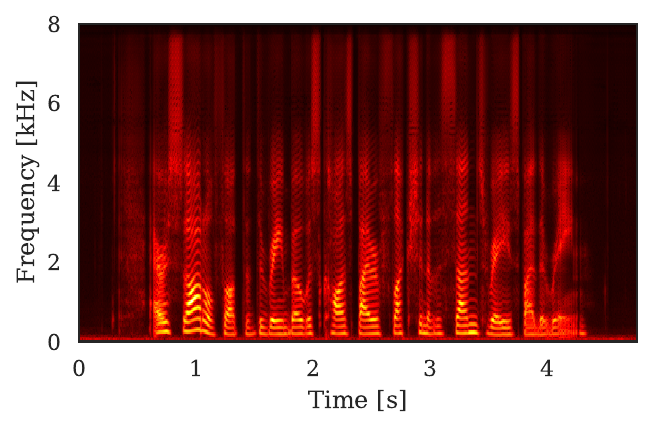}
            \label{fig:gvis_input_c}
            \end{minipage}
    }
    \hfill
    \subfloat[\rmfamily\footnotesize Multi-distortion]{
       \begin{minipage}{0.23\textwidth}
            \centering
            \includegraphics[width=\linewidth]{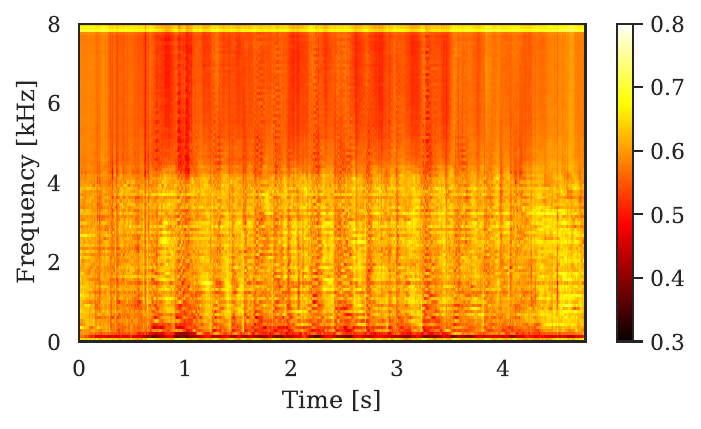}\\
            \includegraphics[width=0.9\linewidth]{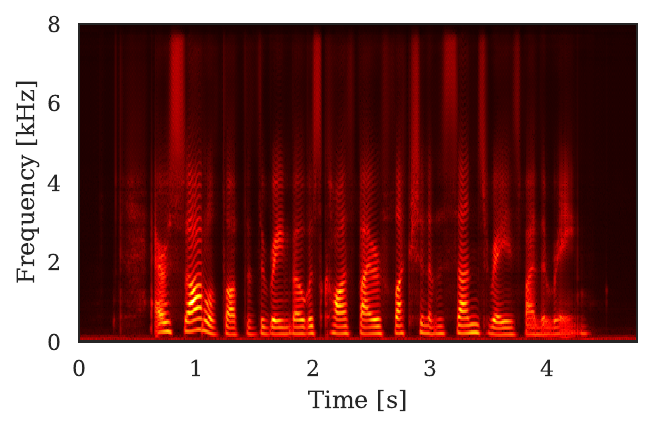}
            \label{fig:gvis_input_d}
            \end{minipage}
    }
   \caption{Visualization of gate maps in input-specific scenarios using the multi-distortion model. Gate maps and output magnitude spectrogram of the multi-distortion model given input of (a) additive noise, (b) reverberation, (c) 8 kHz bandwidth limitation, and (d) all three distortions combined.}
   \vspace{-4mm}
   \label{fig:gvis_input}
\end{figure*}
\Tref{tab:gate_mean_task} summarizes the average gate activations for models trained on individual distortion types. The denoising model exhibits the strongest masking tendency (0.772), followed by the bandwidth extension models at 8 kHz (0.760) and 4 kHz (0.731). The dereverberation model shows the lowest masking ratio (0.693), reflecting increased reliance on mapping. These trends align with the nature of each task. In denoising, where speech remains largely intact and noise is additive, masking effectively suppresses interference. Dereverberation introduces globally diffuse and smeared distortions, necessitating broader reconstruction through mapping. Bandwidth extension displays more nuanced behavior: as input bandwidth narrows, the model increasingly depends on mapping to infer missing high-frequency components, while preserving low-frequency content through masking.

\Fref{fig:gvis_task} visualizes this behavior. Under additive noise and reverberation, the gating pattern remains broadly uniform; however, reverberation yields slightly lower gate values, reflecting a modestly greater reliance on mapping compared to denoising. For bandwidth-limited inputs, the gate map reveals strong frequency dependency: masking dominates in the lower band and gradually diminishes at higher frequencies, where reconstruction is required. This pattern resembles prior approaches that explicitly reuse the low-frequency band~\cite{liu2022neural} or apply residual learning~\cite{wang2021towards, lu2024towards}; however, in our case, the division emerges organically without any frequency-specific architectural design. The model learns to exploit the reliability of the lower band and invoke mapping only where necessary, suggesting that the gating mechanism adaptively modulates its behavior in response to match task requirements.

\subsubsection{Input-Specific Behavior}
We next assess whether the gating mechanism in the multi-distortion model remains adaptive to input characteristics despite being jointly trained. \Tref{tab:gate_mean_input} reports average gate values for test subsets grouped by distortion type. As expected, masking weights decrease with reduced bandwidth, from 0.594 at 8 kHz to 0.569 at 2 kHz, indicating an adaptive shift toward mapping. In contrast, variations in SNR have little effect on gating behavior, showing that masking remains robust across different noise intensities.

\Fref{fig:gvis_input} presents the gate maps and enhanced magnitude spectrograms for both individual and combined distortions. Compared to single-task models, the multi-distortion model produces more complex spatial patterns, likely reflecting its exposure to overlapping distortion types during training. Despite this complexity, distinct gating behaviors remain evident: additive noise and reverberation lead to broadly distributed masking, while bandwidth limitation induces stronger mapping activity in high-frequency regions. Notably, under single bandwidth limitation condition (\Fref{fig:gvis_input_c})—a scenario to which the multi-distortion model was not explicitly exposed during training—the module still generates fine-grained masking peaks that align with harmonic structures. The output spectrograms verify that these gating decisions facilitate effective signal recovery even in such unseen conditions; when the low-frequency band remains clean, the model reconstructs a high-frequency energy distribution that closely approximates the ground-truth reference. In multi-distortion cases (\Fref{fig:gvis_input_d}), however, the model invokes a more intricate combination of masking and mapping to restore corrupted low-frequency regions. While this yields favorable results, the inherent difficulty of low-frequency reconstruction leads to subtle deviations in the high-frequency energy, potentially acting as a primary source of high-frequency artifacts. Overall, these results confirm that the gating module not only differentiates distortion types but also adaptively modulates its strategy based on the spectral characteristics of the input signal.

\subsubsection{Summary and Implications}
The GM module serves as a dynamic controller that adjusts enhancement strategies based on task requirements and input spectral context. Its ability to implicitly distinguish between reliable and missing regions—particularly under bandwidth-limited conditions—demonstrates perceptually aligned behavior learned without explicit priors. This context-aware fusion is central to EDNet’s ability to perform well across varied tasks and overlapping distortions.

\begin{table}[!t]
    \centering
    \caption{Comparison of Parameter counts and FLOPs.}
    \vspace{-2mm}
    \resizebox{0.99\linewidth}{!}{
    \begin{tabular}{l|c|ccc|c}
        \toprule
        \textbf{Method} & \textbf{Param.} & \textbf{Encoder} & \textbf{Core Module} & \textbf{Decoder} & \textbf{Total} \\
        \midrule
        CMGAN\cite{abdulatif2024cmgan} & 1.83M & 16.40G & 29.20G & 17.70G & 63.30G \\
        SEMamba \cite{chao2024investigation}   & 2.25M & 24.36G & 16.00G & 24.60G & 64.96G \\
        MP-SENet \cite{lu2023explicit}  & 2.26M & 16.40G & 52.40G & 17.72G & 86.52G \\
        \midrule
        \textbf{EDNet} & 2.67M &  20.52G & 31.04G & 11.10G & \textbf{62.66G}\\
        \hspace{0.2em}- Mag stream &  2.28M & 16.40G & 27.08G & 8.86G & 52.34G \\
        \hspace{0.2em}- Pha stream & 0.39M & 4.12G & 3.96G & 2.24G & 10.32G\\
        \bottomrule
    \end{tabular}
    }
    \label{tab:flops}
    \vspace{-4mm}
\end{table}
\subsection{Model Complexity Analysis}
In terms of computational cost, EDNet differs from prior DenseNet-based encoder–decoder models in two respects: (i) a dual-stream architecture that processes magnitude and phase separately, and (ii) stacked Gating Mamba (GM) blocks, each comprising two processing modules—masking and mapping. Managing the associated overhead is a primary concern in our model design. In this section, we analyze the design choices intended to address this overhead and their effects, from theoretical and empirical perspectives.

~\Tref{tab:flops} summarizes the parameter counts and FLOPs of EDNet and comparable DenseNet-based systems. While EDNet introduces a modest increase in parameters, it attains the lowest total FLOPs. These efficiency gains arise from three design choices. First, to offset the doubled computation incurred by executing both masking and mapping process, we select the TF-Mamba~\cite{chao2024investigation} module for its superior computational efficiency relative to Transformer-based alternatives. Second, whereas most DenseNet-based baselines employ four processing blocks, EDNet uses three GM blocks. Third, to minimize the overhead introduced by the dual-stream design, we allocate the majority of the capacity to the magnitude stream and make the phase stream deliberately lightweight by reducing its channel width to 32 (0.39M parameters; 10.32G FLOPs). This magnitude-dominant allocation is motivated by our phase-reconstruction analysis in following \Sref{sec:phase_reconstruction}. We find that providing clean magnitude features enables the model to recover phase near ground-truth quality, indicating that accurate magnitude restoration is more critical to overall enhancement performance.

\begin{figure}[t]
    \centering
    \subfloat[\rmfamily\footnotesize Real-time Factor]{
        \includegraphics[width=0.23\textwidth]{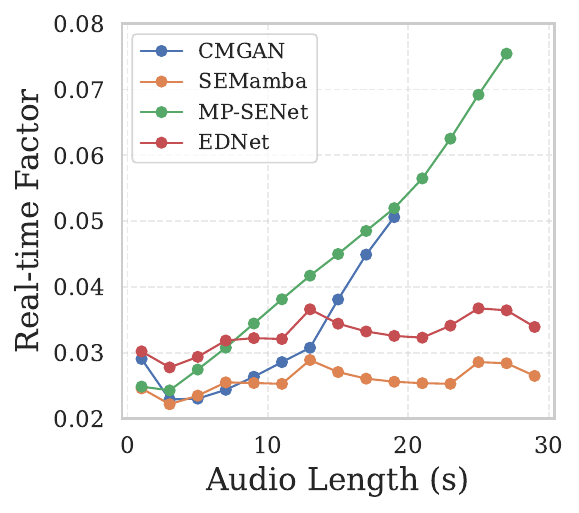}
        \label{fig:rtf}
    }
    \subfloat[\rmfamily\footnotesize Memory Usage]{
        \includegraphics[width=0.23\textwidth]{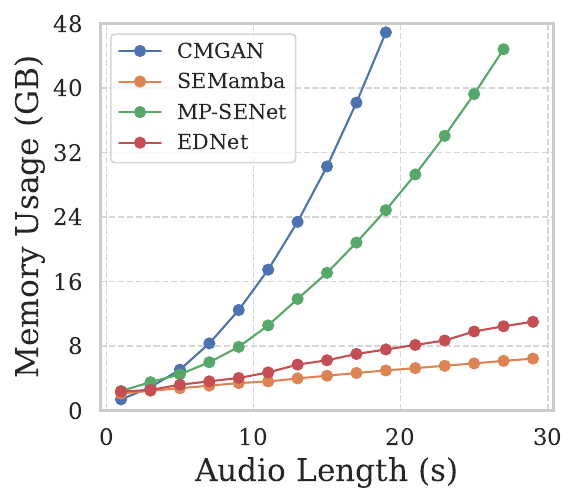}
        \label{fig:mem}
    }
   \caption{Empirical efficiency comparison among CMGAN, SEMamba, MP-SENet and EDNet. (a) Real-time factor (RTF) and (b) GPU memory usage with increasing input duration.}
   \vspace{-2mm}
   \label{fig:rtf_mem}
\end{figure}
In addition to the theoretical complexity analysis, we assess empirical efficiency of EDNet under real-world inference conditions. We measure the average real-time factor (RTF) and GPU memory consumption using 100 audio samples with durations from 1 to 30 seconds in 2-second increments. All experiments are conducted on an AMD EPYC 7543 CPU and a single NVIDIA A6000 GPU (48 GB VRAM). As shown in~\Fref{fig:rtf_mem}, SEMamba~\cite{chao2024investigation} attains the lowest RTF and memory usage, with EDNet following a similar trend. Despite having the lowest FLOPs, EDNet shows a measured RTF that is approximately 0.01 higher and consumes slightly more memory, likely due to the heavier core module and serial magnitude-to-phase reconstruction process.

When compared with Transformer-based counterparts such as CMGAN~\cite{abdulatif2024cmgan} and MP-SENet~\cite{lu2023explicit}, the Mamba-based models (EDNet and SEMamba) show clear advantages on longer sequences, maintaining relatively stable RTF and memory usage across input durations. CMGAN and MP-SENet begin to lag behind EDNet at approximately 9 and 15 seconds, respectively, with both RTF and memory consumption rising sharply as input length increases. While the Mamba-based models keep memory usage under 16 GB even for 30-second audio, their processing limits on the 48 GB A6000 are reached at around 19 and 27 seconds, respectively, underscoring the scalability gap between Transformer- and Mamba-based architectures. These findings are consistent with prior studies showing that Mamba architectures scale more efficiently than Transformer-based designs for long-context processing~\cite{gu2024mambalineartimesequencemodeling, jiang2025speech}.

In summary, key design choices of EDNet enable higher enhancement performance and broader applicability, but they can also increase computational demand. By adopting TF-Mamba, reducing the number of GM blocks, and allocating capacity toward the magnitude stream, we control the resulting overhead so that both theoretical and measured complexities remain comparable to DenseNet-based baselines. Across all tested input lengths, EDNet maintains a real-time factor below 0.04 with practical memory usage, meeting real-time requirements and scaling to long contexts.

\vspace{4mm}
\subsection{Phase Analysis and Impact of PSIT}
\label{sec:psit}
We now turn to the proposed phase shift-invariant training (PSIT) strategy, which is designed to alleviate the rigid supervision imposed by conventional phase loss functions. This section presents a series of analyses examining the theoretical rationale behind PSIT, its influence on perceptual metrics and phase recovery, and its practical effects during training.

\subsubsection{Behavior of Phase Loss under Phase Shifts}
\begin{figure}[t]
   \centering
   \includegraphics[width=0.85\linewidth]{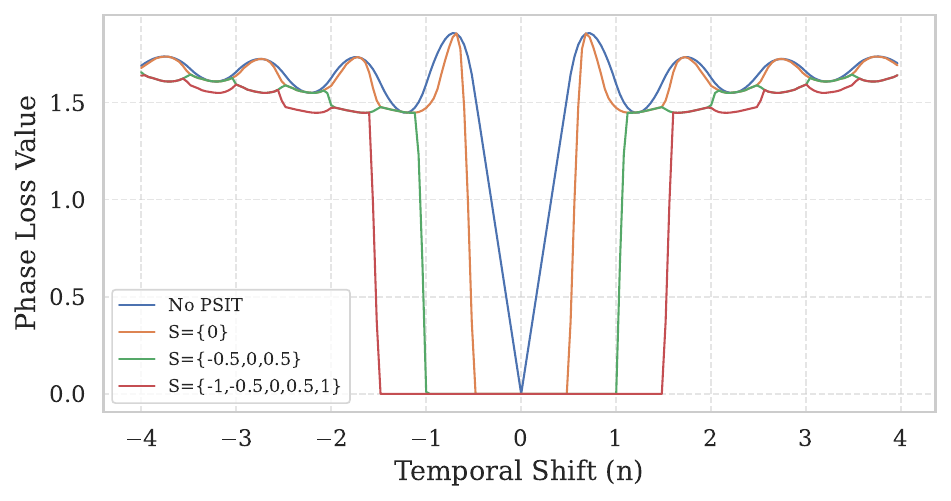}
   \vspace{-2mm}
   \caption{Phase loss with respect to temporal shift $n$. Larger search grids in PSIT lead to wider regions with near-zero phase loss, indicating increased shift tolerance range.}
   \label{fig:shift_phase_loss}
\vspace{-2mm}
\end{figure}

To examine how models receive supervision under phase misalignment, we conduct a controlled experiment using clean utterances from the VoiceBank+DEMAND dataset~\cite{valentini2016investigating}. Synthetic phase shifts are applied to the phase spectrogram using~\Eref{eq:phase_shift}, followed by re-wrapping the phase values into the range $[-\pi,\pi]$ to maintain valid representation. We then compute the loss against the original (unshifted) reference using standard phase loss~\cite{lu2023explicit}, both with and without PSIT, across various search grid configurations.

As shown in~\Fref{fig:shift_phase_loss}, the standard phase loss without PSIT is minimized only when the prediction is perfectly aligned, and it increases sharply with even slight misalignments, exhibiting fluctuations due to phase wrapping. In contrast, PSIT maintains near-zero loss over a wider range, determined by the selected search grid. For instance, a grid of $\{-0.5, 0, 0.5\}$ yields a zero-loss region spanning $[-1, 1]$, based on the validity of linear approximation within $\pm 0.5$ of each grid point. More importantly, beyond this range, PSIT produces a less fluctuating and more consistent loss increase, avoiding the abrupt spikes observed with the standard phase loss. These behavior suggests that PSIT may offer a more stable and interpretable learning signal, even when the predicted phase is shifted yet structurally correct.

\subsubsection{Perceptual Metric Sensitivity to Phase Shifts}
\begin{table}[t]
    \centering
    \caption{Impact of phase shift on objective speech quality metrics.}
    \vspace{-2mm}
    \resizebox{0.99\linewidth}{!}{
    \begin{tabular}{lcccccc}
        \toprule
        \textbf{Phase Shift} & \textbf{Time Shift} & \textbf{PESQ$\uparrow$} & \textbf{CSIG$\uparrow$} & \textbf{CBAK$\uparrow$} & \textbf{COVL$\uparrow$} & \textbf{STOI$\uparrow$} \\
        \midrule
        $n=0$ (GT) & -   & 4.644 & 5.000 & 5.000 & 5.000 & 1.000  \\
        $n=0.33$ & -  & 4.644 & 5.000 & 4.977 & 5.000 & 1.000  \\
        $n=0.66$ & - & 4.643 & 5.000 & 4.769 & 5.000 & 1.000  \\
        \midrule
        $n=1$  & - & 4.643 & 5.000 & 4.627 & 5.000 & 1.000  \\
        $n=1.33$ & -  & 4.642 & 5.000 & 4.541 & 5.000 & 1.000 \\
        $n=1.66$ & -  & 4.641 & 5.000 & 4.466 & 5.000 & 1.000 \\
        \midrule
        $n=1$ & -1 frame  & 4.644 & 5.000 & 4.868 & 5.000 & 1.000  \\
        $n=1.33$ & -1 frame  & 4.643 & 5.000 & 4.686 & 5.000 & 1.000 \\
        $n=1.66$ & -1 frame  & 4.642 & 5.000 & 4.583 & 5.000 & 1.000 \\
        \bottomrule
    \end{tabular}
    }
    \label{tab:shift_metrics}
    \vspace{-5mm}
\end{table}
To assess how phase shifts influence standard perceptual objective metrics, we conduct an analysis using the clean set from the VoiceBank+DEMAND dataset~\cite{valentini2016investigating}. Phase shifts of varying degrees are applied to the clean phase spectrogram in the STFT domain, followed by inverse STFT to reconstruct the waveform. The resulting signals are then evaluated using PESQ, CSIG, COVL, CBAK, and STOI~\cite{hu2007evaluation}, as summarized in~\Tref{tab:shift_metrics}. We observe that PESQ exhibits only slight degradation with increasing phase shift, while CSIG, COVL, and STOI remain virtually unchanged across all shift levels. The only metric that consistently declines is CBAK, which is known to be sensitive to sample-level misalignment and likely interprets the resulting waveform discrepancies caused by a temporal shift as background noise artifacts.

To further investigate this behavior, we select samples with large phase shifts (e.g., $n = 1.33$) and perform a one-frame backward shift in the time domain. This correction partially restores PESQ and CBAK scores, supporting the hypothesis that the degradation is primarily due to misalignment. However, compared to phase-shifted signals with equivalent net offsets (e.g., $n = 0.33$), the time-corrected outputs still show slight residual degradation, likely attributable to windowing functions and quantization error introduced during STFT and ISTFT. These results confirm that moderate phase shifts have negligible impact on most perceptual metrics and that degradations observed in CBAK under shift-tolerant training settings such as PSIT should be interpreted with caution.

\label{sec:phase_reconstruction}
\subsubsection{Phase Reconstruction with and without PSIT}
\begin{figure*}[!t]
   \centering
    \includegraphics[width=0.8\textwidth]{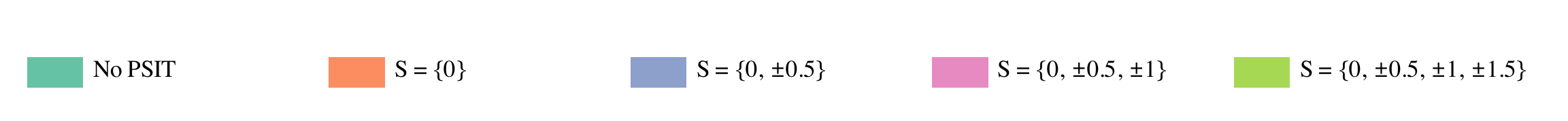}\\
    \vspace{-2mm}
    \subfloat[\rmfamily\footnotesize PESQ]{
        \includegraphics[width=0.27\textwidth]{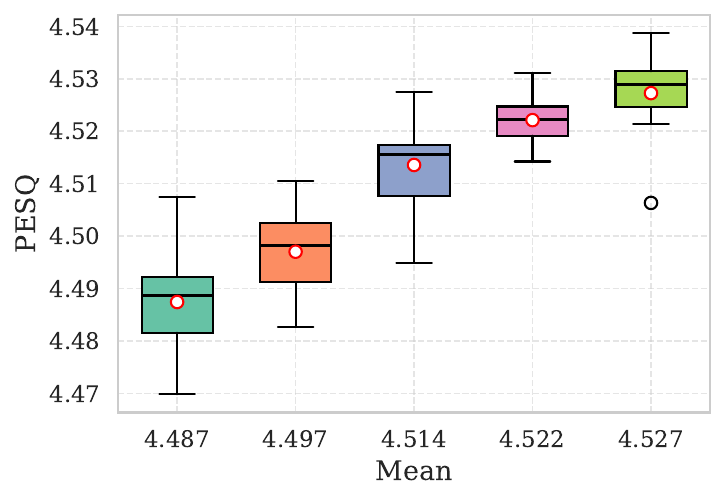}
        \label{fig:pha_recon_pesq}
    }
    \subfloat[\rmfamily\footnotesize CSIG]{
        \includegraphics[width=0.27\textwidth]{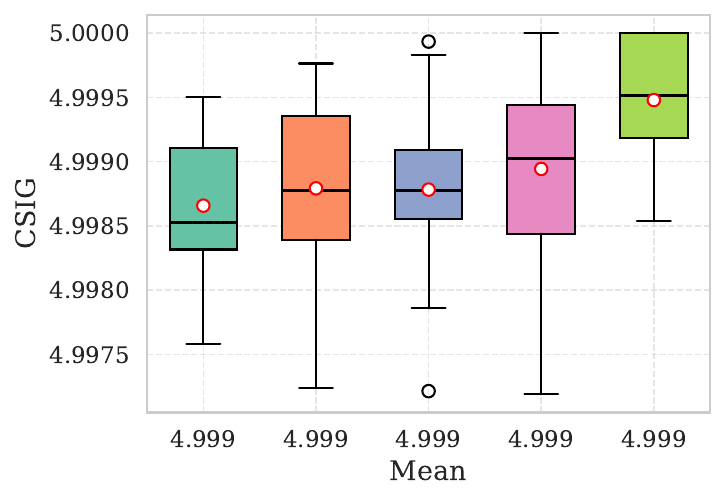}
        \label{fig:pha_recon_csig}
    }
    \subfloat[\rmfamily\footnotesize CBAK]{
        \includegraphics[width=0.27\textwidth]{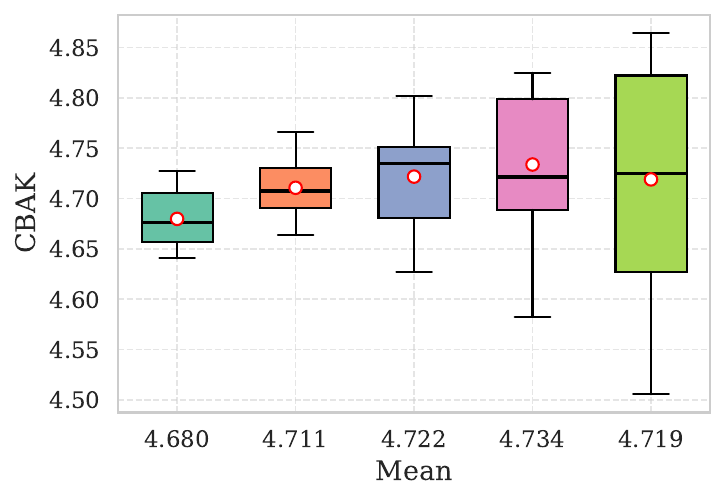}
        \label{fig:pha_recon_cbak}
    }\\
    \subfloat[\rmfamily\footnotesize COVL]{
        \includegraphics[width=0.27\textwidth]{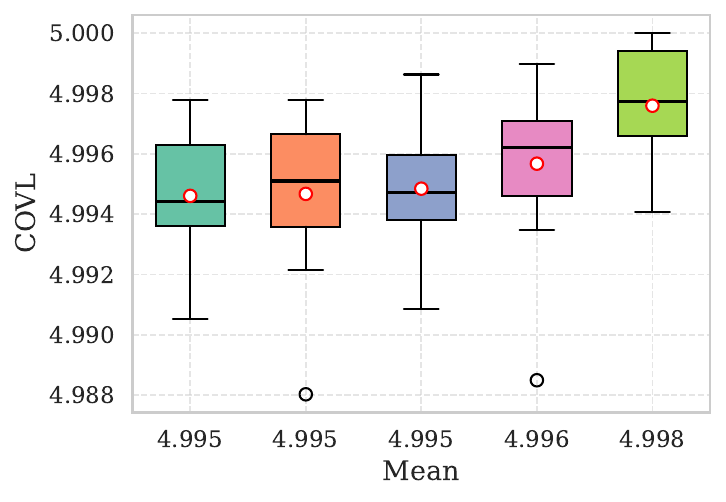}
        \label{fig:pha_recon_covl}
    }
    \subfloat[\rmfamily\footnotesize STOI]{
        \includegraphics[width=0.27\textwidth]{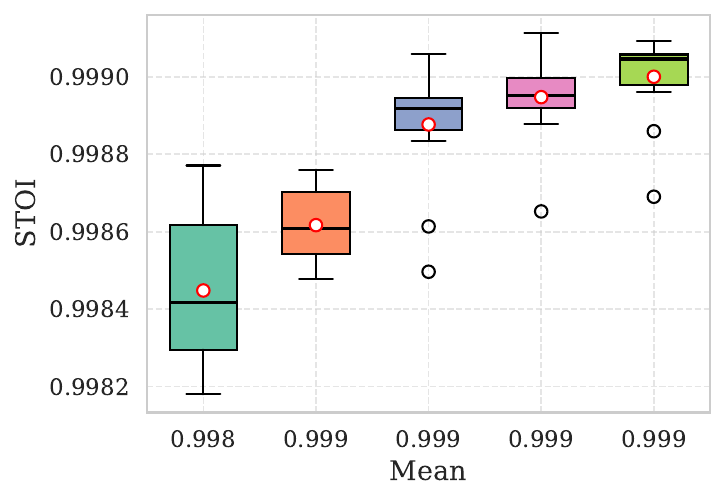}
        \label{fig:pha_recon_stoi}
    }
    \subfloat[\rmfamily\footnotesize PSI-PD]{
        \includegraphics[width=0.27\textwidth]{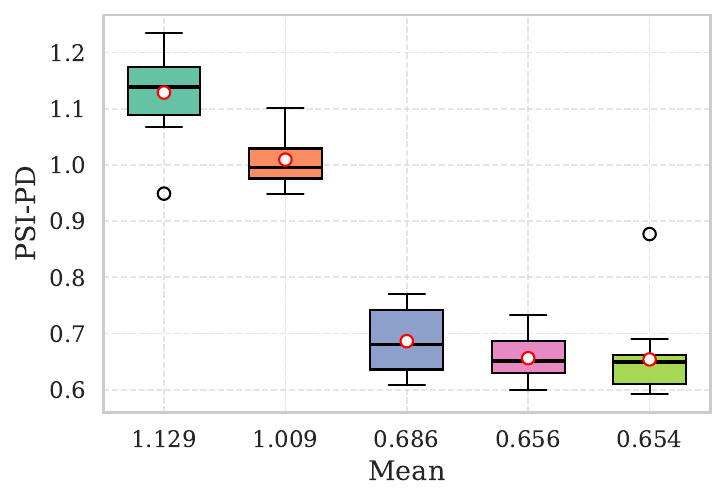}
        \label{fig:pha_recon_psipd}
    }
   \caption{Box plots comparing performance in the phase reconstruction task under different PSIT configurations. Results are shown for without PSIT and with PSIT of varying search grids. Each box height represents the interquartile range (Q1 to Q3), with the median shown as a solid line inside the box, and the mean indicated by a red-edged circle. Whiskers extend to the minimum and maximum values excluding outliers, which are marked with black-edged circles. Mean values are annotated on the x-axis.}
   \vspace{-4mm}
   \label{fig:pha_recon_box}
\end{figure*}

\begin{figure*}[t]

    \centering
    \subfloat[\rmfamily\footnotesize Magnitude Loss]{
        \includegraphics[width=0.22\textwidth]{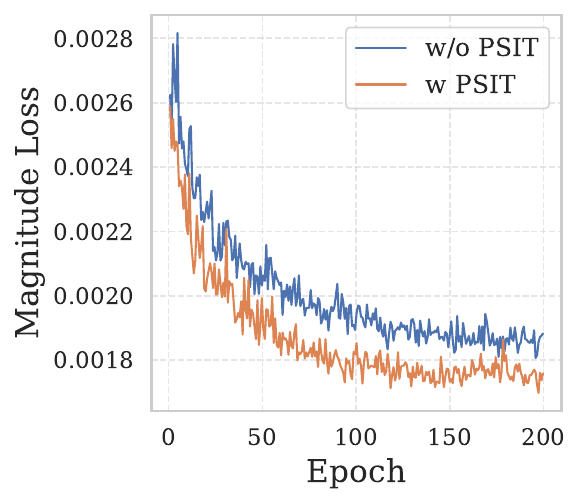}
        \label{fig:vcurve_mag}
    }
    \subfloat[\rmfamily\footnotesize Phase Loss]{
        \includegraphics[width=0.22\textwidth]{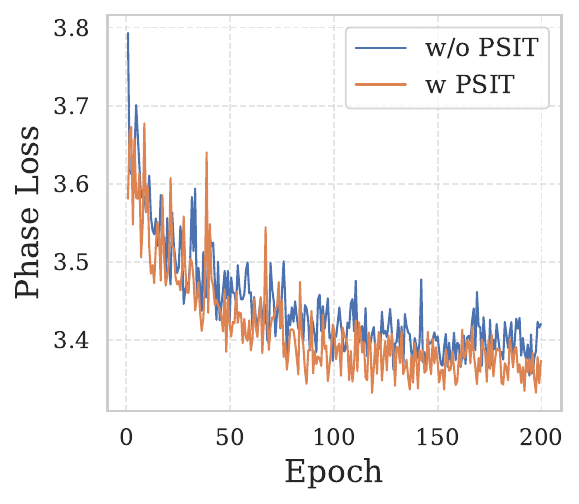}
        \label{fig:vcurve_pha}
    }
    \subfloat[\rmfamily\footnotesize Complex Loss]{
        \includegraphics[width=0.22\textwidth]{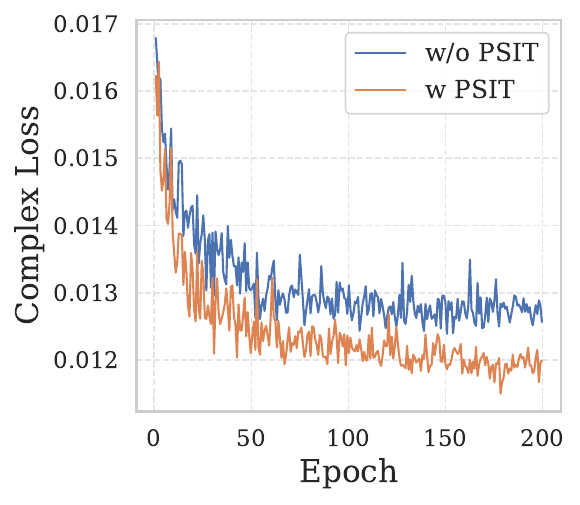}
        \label{fig:vcurve_comp}
    }
    \subfloat[\rmfamily\footnotesize PESQ]{
        \includegraphics[width=0.22\textwidth]{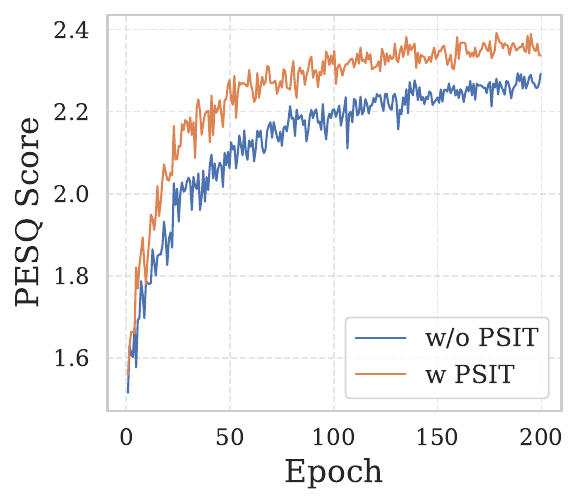}
        \label{fig:vcurve_pesq}
    }
   \caption{Validation curve of (a) Magnitude Loss, (b) Phase Loss, (c) Complex Loss, and (d) PESQ Score. The x-axis represents the training Epoch and the y-axis represents values.}
   \vspace{-4mm}
   \label{fig:vcurve}
\end{figure*}
For assessing the impact of PSIT on phase reconstruction quality, we conduct a controlled experiment where the model is trained to predict phase from clean ground-truth magnitude. This setup isolates the effect of phase prediction by removing confounding factors related to magnitude estimation. We compare the standard phase loss~\cite{lu2023explicit} (w/o PSIT) against four PSIT variants with increasing search grid sizes, each trained 15 times with different random seeds to ensure statistical reliability. 

As shown in~\Fref{fig:pha_recon_box}, PESQ scores improve steadily as the search grid expands, with statistically clear gains emerging from the grid $\{0, \pm 0.5\}$. CSIG, COVL, and STOI remain close to ground-truth levels across all configurations, implying that these metrics are largely saturated under the clean-magnitude setup. While CSIG and COVL show minimal variation, STOI exhibits a statistically clearer upward trend, suggesting that PSIT still provides marginal but repeatable improvements in phase reconstruction. For CBAK, average scores slightly increase with wider PSIT grids, though variance becomes more pronounced—reflected in broader interquartile ranges and more extreme values. This may result from PSIT allowing a wider range of phase predictions that are perceptually valid but not strictly aligned, introducing variability in waveform-level alignment. These findings align with previous observations that CBAK is sensitive to sample-level discrepancies and may over-penalize benign phase deviations. To complement the analysis, we introduce Phase Shift-Invariant Phase Distance (PSI-PD) as an auxiliary evaluation measure. While mathematically equivalent to the standard phase loss with PSIT, PSI-PD employs a denser and wider search grid (step size 0.25, range -20 to 20), allowing it to discount errors that arise purely from phase shifts and to focus instead on the structural accuracy of phase reconstruction. In our experiments, PSI-PD scores decreased consistently with larger PSIT grids, reflecting improved phase reconstruction fidelity and aligning closely with PESQ trends. 

Overall, the results indicate that PSIT, when properly configured, can enhance phase reconstruction performance by relaxing rigid alignment constraints. However, given the trade-off between perceptual gains and increased variability in sample-level sensitive metrics like CBAK, moderate search grids (e.g., $\{0, \pm 0.5, \pm 1\}$) may offer the most balanced configuration.

\subsubsection{Training Dynamics in Multi-distortion Setting}

We analyze the impact of PSIT on training dynamics in the full multi-distortion setting.~\Fref{fig:vcurve} presents validation curves for magnitude loss, phase loss, complex loss, and PESQ score over training epochs. While the phase loss shows only a modest reduction under PSIT, the most noticeable improvements are seen in the magnitude and complex losses, both of which converge more rapidly and reach lower minimum values. PESQ also increases steadily throughout training with PSIT, indicating enhanced perceptual quality as a result of better joint reconstruction of magnitude and phase.

These results suggest that relaxation of strict alignment constraints benefits not only phase reconstruction but also magnitude learning. This may be attributed to the inherently conditional nature of phase prediction: since phase must be estimated in conjunction with magnitude, inaccurate phase supervision—particularly from misaligned targets—can generate incoherent or overly harsh gradients that flow back into the magnitude stream. These noisy signals can disrupt the magnitude learning process, especially in the early training stages when the model is still unstable. By allowing supervision to focus on phase structure rather than exact alignment, PSIT can reduce such interference and enable more coherent gradient flow. 
In this sense, PSIT functions not merely as a phase enhancement tool but also as a training stabilizer for phase-aware speech enhancement frameworks. 

\subsubsection{Summary and Implications}

In summary, our analyses confirm that PSIT effectively fulfills its goal of making phase supervision more tolerant to perceptually irrelevant shifts, both conceptually and empirically. As a training principle, this relaxation improves phase reconstruction quality and also facilitates more efficient and stable optimization of the entire model, including the magnitude stream. This flexibility, however, introduces minor considerations: wider search grids can increase variance in alignment-sensitive metrics such as CBAK, and a small SRMR drop is observable in the real-data dereverberation task. These effects reflect an inherent trade-off—relaxing phase supervision improves overall perceptual quality but can produce temporally shifted output waveform to which certain metrics are sensitive. Therefore, in practical use, the search grid size should be chosen based on the task goal and application, as it directly controls the amount of output shift that PSIT permits. Nonetheless, primary strength of PSIT is its practicality. It operates as a rule-based strategy that introduces only negligible computational overhead during training. It therefore offers a simple and lightweight addition to existing frameworks, yielding measurable gains in performance and efficiency without requiring architectural modifications.

\section{Limitations}
Although this study demonstrates the effectiveness of a unified framework across multiple speech enhancement tasks, several limitations remain. The work primarily focuses on architectural flexibility and task coverage, without including cross-domain evaluations—an important direction for future work aimed at broader generalization. Moreover, while three major distortion types and their combinations are addressed, real-world speech is often affected by a wider range of degradations not covered here, such as packet loss and codec artifacts. Extending the framework to accommodate a broader spectrum of distortions and their simultaneous occurrence represents a valuable avenue for future exploration. Lastly, despite strong denoising results, performance in this task remains comparatively lower than in others, indicating room for further improvement.
\section{Conclusion}

This paper proposes EDNet, a versatile speech enhancement framework designed to address diverse distortion types through two key components: a gated integration of masking and mapping operations, and a phase-shift-invariant training (PSIT) objective. The GM module enables the adaptive fusion of complementary enhancement strategies, while PSIT introduces robustness to temporal misalignment in phase supervision. Together, these components form a flexible and generalizable architecture for speech enhancement. We validate the proposed design through comprehensive evaluations on single- and multi-distortion tasks, where EDNet achieved state-of-the-art or comparable performance without task-specific architectural adaptation. Ablation studies and gating behavior analyses confirm that the GM module not only improves performance but also allows the model to learn task- and input-specific control over its enhancement strategy. In addition, PSIT improves phase reconstruction quality, facilitates learning dynamics, and improve overall model performance by softening the effects of strict alignment constraints without an overhead during training.
These results highlight that integrating structural adaptability with shift-tolerant supervision offers a promising path toward building versatile speech enhancement systems suited for real-world deployment
\section{Acknowledgement}
This work was supported by the Institute of Information \& communications Technology Planning \& Evaluation (IITP) through a grant funded by the Korean government (MSIT; RS-2025-02215122, 50\%) and by the IITP–ITRC program funded by the Korean government (MSIT; IITP-2026-RS-2023-00259991, 50\%).

\bibliographystyle{IEEEtran}
\bibliography{shortstrings, mybib}
\section{Biography Section}
\begin{IEEEbiographynophoto}{Doyeop Kwak}
is a Ph.D. student in Electrical Engineering at the Korea
Advanced Institute of Science and Technology. His main research interests include speech processing and dataset development.
\end{IEEEbiographynophoto}
\begin{IEEEbiographynophoto}{Youngjoon Jang}
is a postdoctoal research fellow at the Korea Advanced Institute of Science and Technology. He received the Ph.D. degree in Electrical Engineering from Korea Advanced Institute of Science and Technology. His research interests include video understanding and multimodal learning.
\end{IEEEbiographynophoto}
\begin{IEEEbiographynophoto}{Seongyu Kim}
is a Ph.D. student in Electrical Engineering at the Korea Advanced Institute of Science and Technology. His research interests include audio-visual alignment and multimodal learning.
\end{IEEEbiographynophoto}
\begin{IEEEbiographynophoto}{Joon Son Chung}
is an associate professor at the Korea Advanced Institute of Science and Technology, where he is directing research in speech processing, computer vision and machine learning. He received the D.Phil. in Engineering Science from the University of Oxford.
\end{IEEEbiographynophoto}

\vfill

\end{document}